\documentclass{aa}

\usepackage[varg]{txfonts}
\usepackage{graphicx}

\usepackage{xcolor}
\usepackage{mathrsfs}
\usepackage{booktabs}
\usepackage{tabularx}
\usepackage{multicol}
\usepackage{graphicx}
\usepackage{subcaption}
\usepackage[subfigure]{tocloft}

\RequirePackage{nameref}

\begin{document}

\title{Spatial Variations of Polarized Synchrotron Emission in the QUIJOTE MFI Data using Neural Networks}
\titlerunning{Synchrotron and QUIJOTE data with CNN}
\authorrunning{J.M. Casas et al.}

    \author{J. M. Casas \inst{1,2,3,4},
    L. Bonavera \inst{5,2},
    J. Gonz{\'a}lez-Nuevo \inst{5,2},
    J. A. Rubiño-Mart{\'i}n \inst{3,4},
    R. T. G{\'e}nova-Santos \inst{3,4},
    R. B. Barreiro \inst{6},
    M. M. Cueli \inst{1,2,7,8},
    D. Crespo \inst{5,2},
    R. Fern{\'a}ndez-Fern{\'a}ndez \inst{5,2},
    J. A. Cano \inst{5,2}
    }
    
  \institute{
  $^1$Universidad de Oviedo, Departamento de Matem\'{a}ticas, C. Federico Garc\'{i}a Lorca 18, 33007 Oviedo, Spain\\
  $^2$Instituto Universitario de Ciencias y Tecnolog\'{i}as Espaciales de Asturias (ICTEA), C. Independencia 13, 33004 Oviedo, Spain\\
  $^3$Instituto de Astrofísica de Canarias, E-38205 La Laguna, Tenerife, Spain\\
  $^4$Universidad de La Laguna, Departamento de Astrofísica, E-38206 La Laguna, Tenerife, Spain\\
  $^5$Universidad de Oviedo, Departamento de F\'{i}sica, C. Federico Garc\'{i}a Lorca 18, 33007 Oviedo, Spain\\
  $^6$Instituto de Física de Cantabria (IFCA), CSIC-Univ. de Cantabria, Avenida los Castros, s/n, E-39005 Santander, Spain\\
  $^7$SISSA, Via Bonomea 265, 34136 Trieste, Italy\\
  $^8$IFPU - Institute for fundamental physics of the Universe, Via Beirut 2, 34014 Trieste, Italy\\         
}
\date{}
\abstract
{Polarized synchrotron emission from ultra-relativistic electrons spiraling the Galactic magnetic field has become one of the most relevant emissions in the Interstellar medium these last years due to the improvement in the quality of low-frequency observations. One of the recent experiments designed to explore this emission is the QUIJOTE experiment.

In this work, we aim to study the spatial variations of the synchrotron emission in the QUIJOTE MFI data, by dividing the sky into physically separated regions. For such task, we firstly use a novel component separation method based on artificial neural networks to clean the synchrotron maps. After training the network with simulations, we fit both $EE$ and $BB$ spectra by assuming a power-law model. Then, we give estimations for the index $\alpha_{S}$, the amplitude, and the ratio between $B$ and $E$ amplitudes. 

When analyzing the real data, we found a clear spatial variation of the synchrotron properties along the sky at 11 GHz, consistent with previous analyses, obtaining a steeper index in the Galactic plane of $\alpha_{S}^{EE} = -3.10 \pm 0.30$ and $\alpha_{S}^{BB} = -3.10 \pm 0.28$ and a flatter one at high Galactic latitudes of $\alpha_{S}^{EE} = -3.05 \pm 0.16$ and $\alpha_{S}^{B} = -2.98 \pm 0.23$. We found average values at all sky of $\alpha_{S}^{EE} = -3.04 \pm 0.18$ and $\alpha_{S}^{BB} = -3.00 \pm 0.26$. Furthermore, after obtaining an average value of $A_{S}^{EE} = 3.31 \pm 0.17$ $\mu K^{2}$ and $A_{S}^{BB} = 0.93 \pm 0.04$ $\mu K^{2}$, we estimate a ratio between $B$ and $E$ amplitudes of $A_{S}^{BB}/A_{S}^{EE} = 0.28 \pm 0.06$. 

Based on the results we conclude that, although neural networks seem to be valuable methods to apply on real ISM observations and in future QUIJOTE MFI2 data, combined analyses with \textit{Planck}, WMAP and/or CBASS data are mandatory to reduce the noise contamination from QUIJOTE estimated maps and then improve the accuracy of the estimations.
}

\maketitle

\section{Introduction}
\label{sec:introduction}

The interstellar medium (ISM) is a complex physical system where several emissions undergo abrupt state changes when interacting with each other and with the Galactic magnetic field \citep{DRA11}, displaying significant variability across all scales. To characterize this complexity, phenomenological models are developed based on observations from ground-based and satellite experiments, as well as accurately inferring the physical behavior of these emissions at scales not yet observed \citep{PANEX}.

In recent decades, several experiments have helped constrain many properties of the ISM, especially in the microwave band. In particular, observations from the Planck satellite \citep{PLA_18_I} have provided unprecedented, detailed maps on both the temperature and polarization emissions. {The \textit{Planck} mission released in the last decade} detailed temperature and polarization maps of interstellar dust emission (\citealt{PLA_13_XI}, \citealt{PLA_15_XIX}), which have led to improvements in theoretical models \citep{HEN21}. However, many properties of synchrotron emission produced by ultra-relativistic electrons spiraling along the Galactic magnetic field remain poorly constrained, especially due to the lower sensitivity of the instruments measuring the lowest frequencies of the microwave spectra, although efforts have been made recently by the BeyondPlanck and COSMOGLOBE Collaborations to published refined maps of the emission \citep{COSMOGLOBE}.

Synchrotron emission is generally assumed to be accurately characterized by a power-law spectral model, provided that the underlying cosmic-ray electron energy distribution likewise exhibits power-law behavior, and that additional effects such as synchrotron self-absorption and the presence of multiple emitting components along the line of sight exert a negligible influence within the microwave regime. Under these conditions, synchrotron emission can, in principle, be modeled as
\begin{equation}
    X_{\nu} \hspace{1pt} = \hspace{1pt} A \hspace{1pt} \left( \hspace{1pt} \frac{\nu}{\nu_{S}} \hspace{1pt} \right) ^{\hspace{1pt}\beta_{S}},
\label{eq:power_law}
\end{equation}
where $X_{\nu}$ represents each $Q$ and $U$ Stokes parameter maps, $A_{S}$ is the brightness temperature at the pivot frequency $\nu_{S} = 30$\,GHz and $\beta_{S}$ is the spectral index, which is assumed to be equal for both $Q$ and $U$ maps.
Moreover, the synchrotron $EE$ and $BB$ power spectra can be fitted over a given multipole range using the parametrization described in \cite{KRA18}:
\begin{equation}
    C_{l}^{XX} \hspace{1pt} = \hspace{1pt} A_{XX} \hspace{1pt} \left( \hspace{1pt} \frac{l}{80} \hspace{1pt} \right) ^{\hspace{1pt}\alpha_{XX}},
\label{eq:power_law_parametrization}
\end{equation}
where $X\in \{ E, B \}$, $A_{XX}$ is the amplitude of the spectrum at the pivot multipole $l=80$, and the index $\alpha_{XX}$ is the slope of the multipole dependence.

Early measurements from WMAP (e.g., \citealt{PAG07}; \citealt{GOL11}) provided large-scale maps of synchrotron polarization and inferred a typical spectral index in polarization of $\beta_{S}\approx-3.0$ with significant spatial variation, especially near the Galactic plane and along filamentary structures. After that, the Planck Low Frequency Instrument (LFI) extended this analysis with improved sensitivity and resolution, giving an average index in polarization of $\beta_{S}\approx-3.1 \pm 0.1$, in combination with WMAP \citep{PLA_18_IV}.

More recent analyses using the same observations alone or in combination with other surveys have better constrained the synchrotron parameters. Particularly, \cite{FUS14} have exploited the WMAP 23 and 33 GHz frequency channels using two parallel methodologies: T–T plots and maximum likelihood estimation across 24 sky regions physically different between them, obtaining a more constrained spatial variation in the synchrotron index, with an average of $\beta_{S} = -2.98 \pm 0.01$ at the Galactic plane and $\beta_{S} = -3.12 \pm 0.04$ at high Galactic latitudes, and with an average value at all sky of $\beta_{S} = -2.99 \pm 0.01$. Moreover, \cite{KRA18} used the S-Band Polarization All Sky Survey (SPASS) data in the range of 2.3 and 33 GHz with several latitude cuts, finding $\alpha_{S}^{E} \approx \alpha_{S}^{B} = -2.59 \pm 0.01$, $A_{S}^{E} = 2.2 \pm 0.2 \hspace{1pt} \mu K^{2}$, $A_{S}^{B} = 1.9 \pm 0.2 \hspace{1pt} \mu K^{2}$ and $A_{S}^{B}/A_{S}^{E} = 0.87 \pm 0.02$ for $|b|>20^{\circ}$ (30\% of the sky), and an average constant value at all sky of $\beta_{S} = -3.22 \pm 0.08$.

After that, \cite{FUS21} joined this data with WMAP 23 GHz one on the same regions described in \cite{FUS14}, and by using a T-T plot methodology, confirming the past analyses concerning the synchrotron spatial variability along the sky, obtaining an index of $\beta_{S} = -2.98 \pm 0.01$ at the Galactic plane and $\beta_{S} = -3.12 \pm 0.04$ at high Galactic latitudes, and with an average value of $\beta_{S} = -3.24 \pm 0.01$ taking into account the Faraday correction.

More recent analyses, such the ones from \cite{MAR22}, \cite{WEI22} and \cite{COSMOGLOBE} have reached similar results. In the first case, they combined WMAP 23 GHz channel with 30 GHz \textit{Planck} one and fitted both $EE$ and $BB$ spectra by assuming a power-law model as in Equations \eqref{eq:power_law} and \eqref{eq:power_law_parametrization} for a set of six sky regions covering from 30 to 94\% of the sky, and finding $\alpha_{S}^{E} = -2.95 \pm 0.04$ and $\alpha_{S}^{B} = -2.85 \pm 0.14$, with a ratio between $B$ and $E$ amplitudes of $r = A_{S}^{B}/A_{S}^{E} = 0.22 \pm 0.02$ for the 94\% case. In the second analysis, they also found a clear spatial variability on synchrotron using SPASS, WMAP and Planck, reaching a synchrotron spectral index in the range $-3.4 \lesssim \beta_{S} \lesssim -3.$ Finally, \cite{COSMOGLOBE} analysis, which is a reprocessed set of maps produced by the COSMOGLOBE Collaboration from all WMAP and \textit{Planck} LFI frequency ones, led to significantly lower instrumental systematics than the legacy products from each experiment. They thus found a spectral index $\beta_{S} = -3.07 \pm 0.07$ at 30 GHz, and a mean ratio between $B$ and $E$ amplitudes of $r = A_{S}^{B}/A_{S}^{E} = 0.39 \pm 0.02$.

All these analyses have confirmed the importance of having data from several experiments with different sensitivities and coverage to be joint together, especially in the low-frequency range. This is precisely the main idea behind the QUIJOTE (Q-U-I JOint Tenerife Experiment) experiment \citep{RUB23}, which have observed the northern sky between 2012 and 2018, and that it is now re-observing it with a more precise instrument called MFI2 \citep{HOY22}. 

QUIJOTE MFI maps have allowed to give more constrains about the synchrotron parameters, for example by fitting the $EE$ and $BB$ power spectra with different Galactic cuts \citep{RUB23}, obtaining $\alpha_{S}^{E} = -3.00 \pm 0.16$ and $\alpha_{S}^{B} = -3.08 \pm 0.42$, and a ratio between $B$ and $E$ amplitudes of $r = A_{S}^{B}/A_{S}^{E} = 0.26 \pm 0.07$, and also by running a parametric component separation method \citep{dlHOZ23}, reaching an average spectral index of $\beta_{S} = -3.08 \pm 0.13$ while \cite{ADA26}, with a semi-blind method, have recently obtained an inverse-variance-weighted mean spectral index of $\beta_{S} = -3.11 \pm 0.21$ using the same data.

In this work, we aim to extend the findings of these two approaches by investigating more about the spatial variability of the polarized synchrotron emission in the public QUIJOTE MFI observations with the use of a novel component separation method based on artificial neural networks, initially presented in \cite{CAS22b}, for separating the cosmic microwave background (CMB) in temperature maps, and thus extended to polarization CMB maps in \cite{CAS25}. The main motivation to use this kind of methodology is the ability of neural networks to find non-trivial patterns in non-linear data, which is precisely the ones governing the emissions of the microwave polarization sky. Section \ref{sec:data} describes the data and realistic simulations we use in this work. Section \ref{sec:methodology} summarizes the methodology. Section \ref{sec:results} presents the results when applying the network to both realistic simulations and observations and Section \ref{sec:conclusions} shows our conclusions. Appendix \ref{sec:additional_tests} shows additional methodological tests, Appendix \ref{sec:model} presents a summary of the theoretical framework of our component separation method and Appendix \ref{sec:full_sky_maps} has the results represented on full-sky maps.

\section{Data}
\label{sec:data}

\subsection{QUIJOTE MFI observations}
\label{sec:data_observations}

\begin{figure*}[t]
\centering
\minipage{0.5\textwidth}
\includegraphics[width=\linewidth]{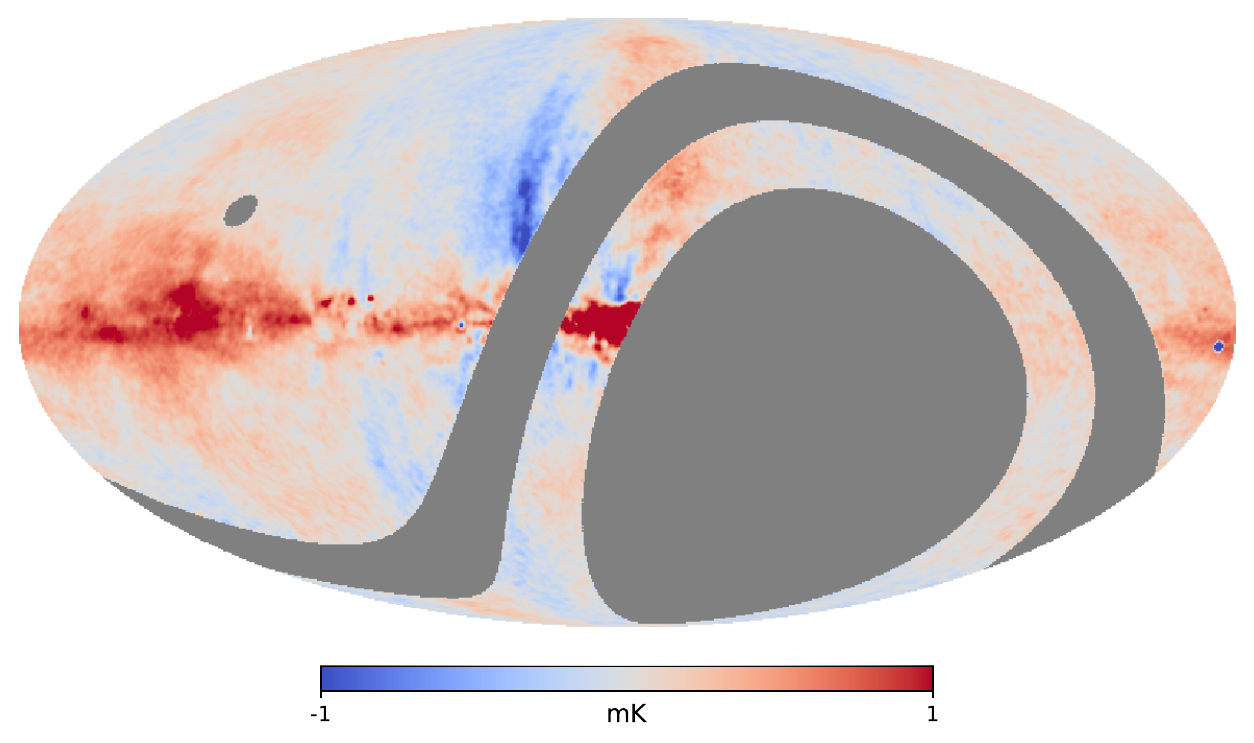}
\endminipage\hfill
\minipage{0.5\textwidth}%
  \includegraphics[width=\linewidth]{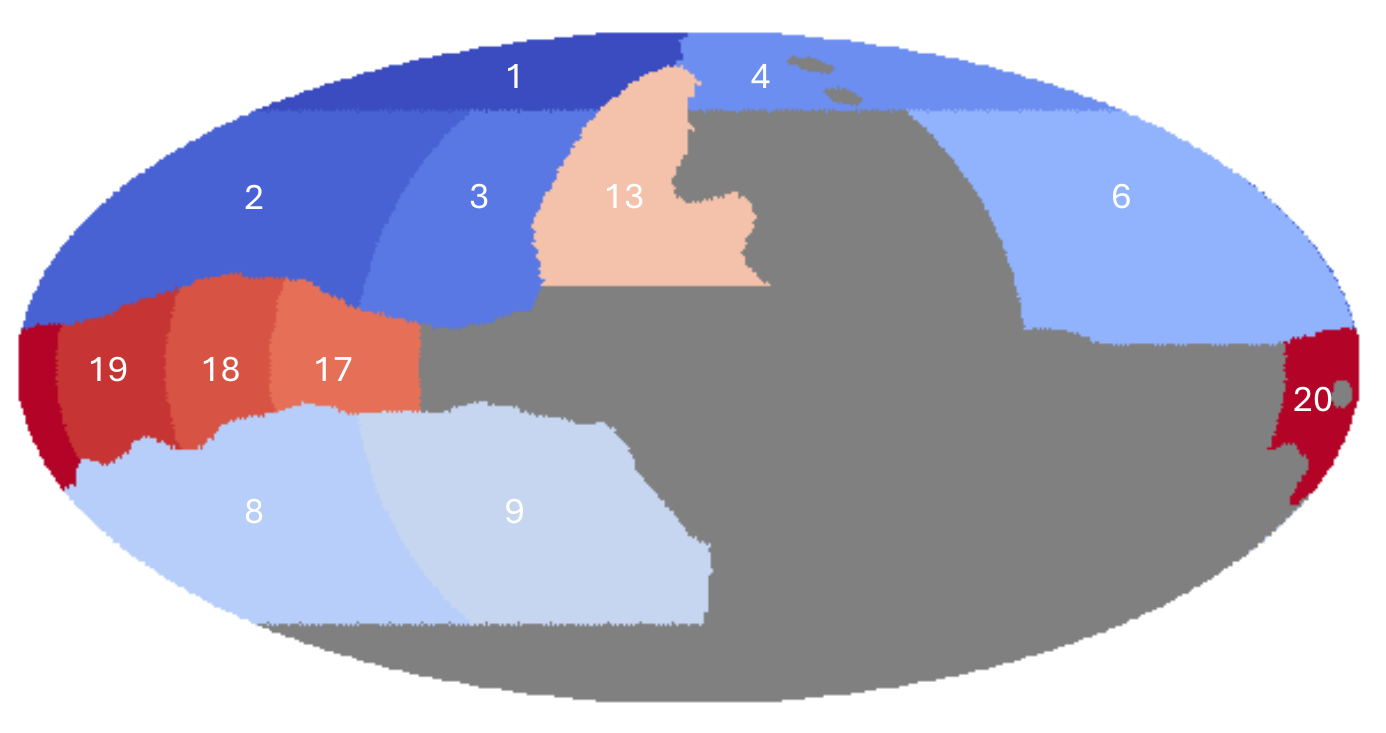}
\endminipage
\caption{Mollview projection of the data used in this work. Left panel: QUIJOTE $Q$ MFI sky at 11 GHz. Right panel: polarized synchrotron regions defined by \cite{FUS14}. It is shown in grey the regions that cannot be used in this work due to QUIJOTE coverage and patch sizes.}
\label{Fig:data}
\end{figure*}

The QUIJOTE (Q-U-I JOint Tenerife Experiment, \cite{RUB23}) project is a ground-based polarimetric experiment focused on studying the polarization of the cosmic microwave background (CMB) as well as various Galactic and extragalactic processes\footnote{https://research.iac.es/proyecto/quijote/}. It operates in the 10-40 GHz frequency range and mainly targets large angular scales. The experiment is situated at the Teide Observatory in Tenerife, at an altitude of roughly 2400 meters, where meteorological conditions are highly stable for observing the sky.

QUIJOTE comprises two telescopes outfitted with three instruments: the Multi-Frequency Instrument (MFI), the Thirty-GHz Instrument (TGI), and the Forty-GHz Instrument (FGI), which observe in the 10-20 GHz, 26-36 GHz, and 39-49 GHz bands, respectively.

The MFI operated from November 2012 until October 2018, during which it carried out two types of surveys: (i) a wide-area Galactic survey covering the sky visible from Tenerife at elevations above 30$^{\circ}$, and (ii) a deep cosmological survey spanning about 3000 square degrees across three separate regions in the northern hemisphere. 
These low-frequency observations are essential for improving our understanding of polarized foreground emission, particularly synchrotron radiation. In combination with \textit{WMAP} and \textit{Planck}, they provide valuable information because the synchrotron signal increases strongly towards lower frequencies, so that, despite its lower absolute sensitivity, QUIJOTE provides a higher signal-to-noise ratio for synchrotron emission.

In this work, we use the MFI observations at 11, 13, 17 and 19 GHz, which are publicly available. The 11 GHz $Q$ sky is shown in Figure \ref{Fig:data}, left panel.

\subsection{QUIJOTE MFI simulations}
\label{sec:data_simulations}

\begin{table}
  \caption{Central positions in longitude and latitude ([lon, lat]) in degrees of the square patches into we divide the sky in this work. Each patch covers a total of 29.44 square degrees of the sky. The regions are the ones defined in \citealt{FUS14} and displayed in Figure \ref{Fig:data}.}
  \label{tab:table_regions}
  \centering
  \begin{tabular}{cc} 
    \hline
    Region (R) & Patch position  \\
    1 & [60$^{\circ}$, 75$^{\circ}$] \\
    2 & [160$^{\circ}$, 45$^{\circ}$] \\
    3 & [60$^{\circ}$, 45$^{\circ}$] \\
    4 & [340$^{\circ}$, 90$^{\circ}$] \\
    6 & [200$^{\circ}$, 45$^{\circ}$] \\
    8 & [120$^{\circ}$, -30$^{\circ}$] \\
    9 & [275$^{\circ}$, -75$^{\circ}$] \\
    13 & [45$^{\circ}$, 45$^{\circ}$] \\
    17 & [90$^{\circ}$, 0$^{\circ}$] \\
    18 & [135$^{\circ}$, 0$^{\circ}$] \\
    19 & [165$^{\circ}$, 0$^{\circ}$] \\
    20 & [225$^{\circ}$, 0$^{\circ}$] \\
    High Galactic Latitudes (HGL) & R1 to R13 \\
    Galactic Plane (GP) & R17 to R20 \\
    \hline
  \end{tabular}
\end{table}

\begin{figure*}[ht]
    \centering
    \includegraphics[width=13.5cm, height=4cm]{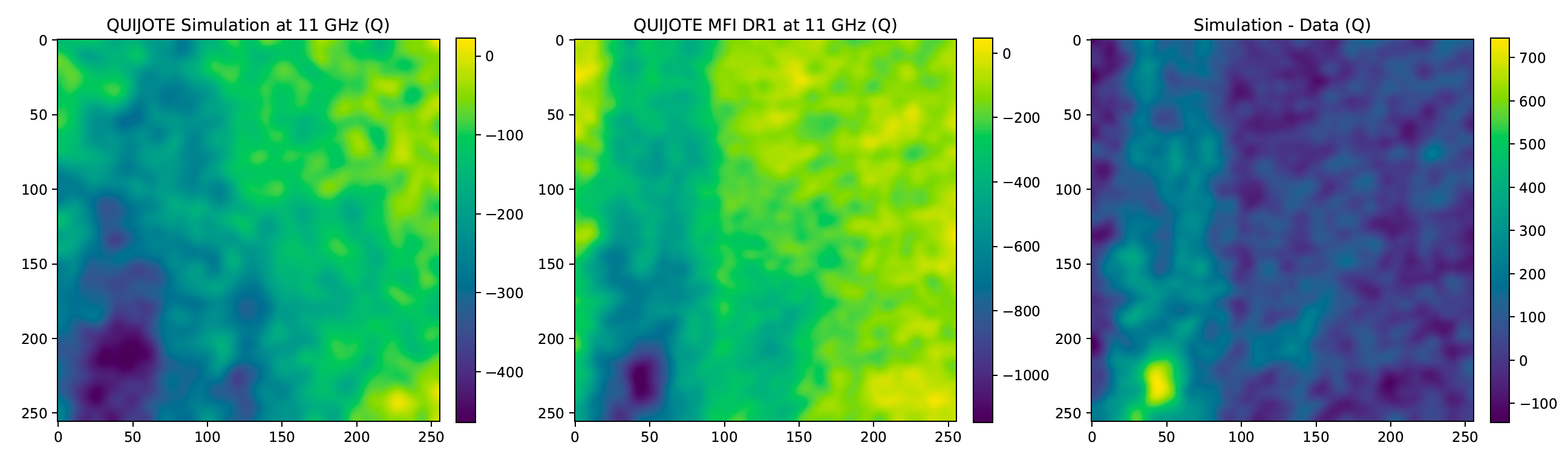}
    \includegraphics[width=13.5cm, height=4cm]{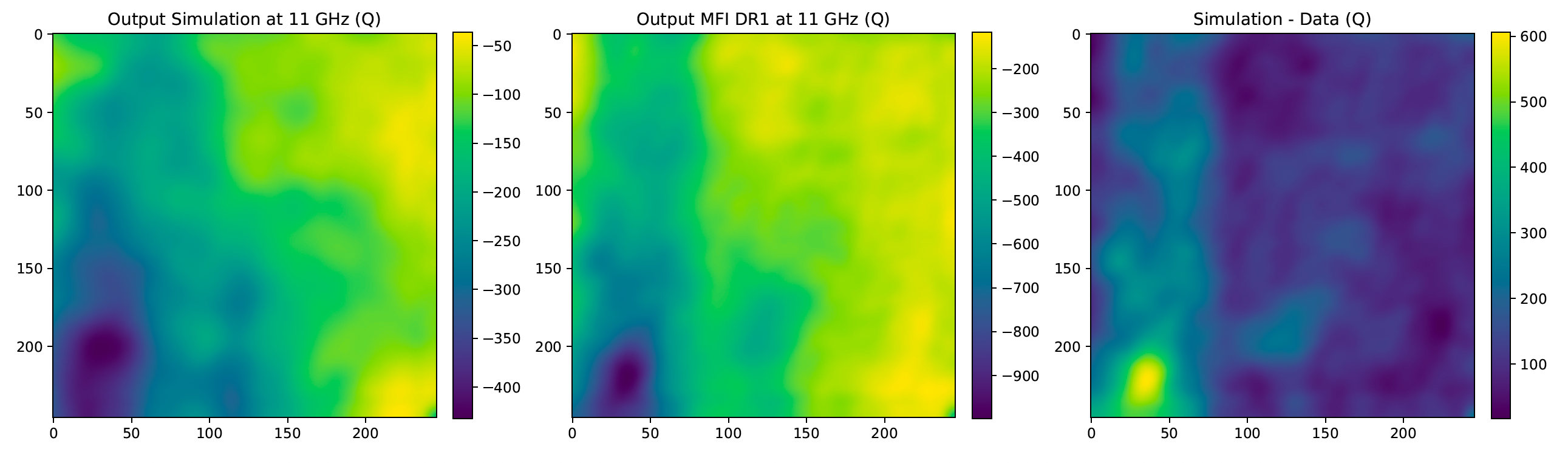}
    \includegraphics[width=13.5cm, height=4cm]{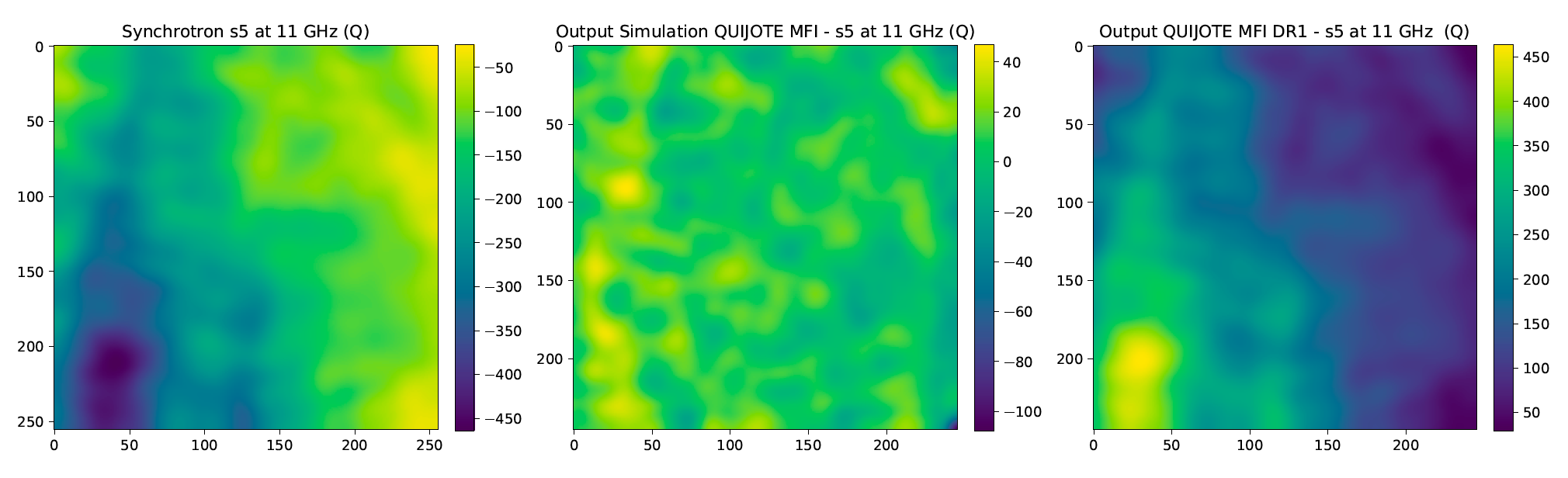}
    \caption{Example of the kind of data we are using in this work, which represents in this case a patch of the $Q$ sky as seen by QUIJOTE MFI in the North Polar Spur (sky Region 13) at the position [lon, lat] = [45$^{\circ}$, 45$^{\circ}$] for, from top to bottom and left to right: our simulations, the QUIJOTE MFI observations at 11 GHz, the difference between them, the synchrotron output patch predicted by the neural network, the output when testing the real QUIJOTE MFI DR1 observations, the difference between these two outputs, the synchrotron s5 model at this sky patch and then the difference between the outputs from the network and the synchrotron model when using both simulations and observations, respectively. Colorbars show the $\mu K_{CMB}$ units for each patch.}
    \label{fig:patches}
\end{figure*}

The simulations used in this work include both Galactic and extragalactic emissions at the frequencies covered by QUIJOTE, as well as the CMB and instrumental noise consistent with QUIJOTE sensitivity levels. The CMB simulation was obtained from the Planck Legacy Archive website (PLA\footnote{http://pla.esac.esa.int/pla/\#home}), as a lensed realization developed from the published cosmological parameters. On the other hand, Galactic dust emission was modeled using the d10 model from the Python Sky Model (PySM, \citep{THO17}), while Galactic synchrotron emission follows the s5 model, also from PySM. This synchrotron model has the particularity that it has no curvature although having spatial variability of the spectral index along the sky inside WMAP and \textit{Planck} observational bounds\footnote{The reader is encouraged to follow the PySM documentation in order to find more information about this: https://pysm3.readthedocs.io/en/latest/models.html}. In this work, we therefore assume that the synchrotron emission adheres to a power-law model; however, a consistency check employing a curvature-based model is presented in the Appendix \ref{sec:curvature_models}.

Anomalous microwave emission and radio point sources were added for completeness using, for the first, the a2 model from the PySM, and for the second the software CORRSKY \citep{GN05} and the model by \cite{TUC11} in intensity, with a final extrapolation to polarization by using the estimated parameters by \cite{Bon17a} at low microwave frequencies.

As reported in \cite{RUB23}, the calibration uncertainty in the QUIJOTE-MFI maps is approximately 5\%, predominantly arising from uncertainties in the modelling of the calibrators. The internal calibration uncertainty of the QUIJOTE-MFI intensity channels is on the order of 1\%. In polarization, this internal uncertainty is chiefly driven by the polarization efficiency, which is approximately between 3 and 4\%. In this work, we incorporate this contribution into the amplitude estimates presented in Table \ref{tab:table_observations} by combining it in quadrature with the remaining sources of uncertainty.

As discussed in \cite{RUB23}, the co-addition of all data during the map-making process of the QUIJOTE-MFI polarization maps introduced residual RFI signals at fixed azimuthal positions, which subsequently projected onto the maps as stripes of constant declination. These artifacts were mitigated through the application of a declination-dependent function (hereafter FDEC), effectively equivalent to filtering the maps by removing the zero mode along lines of constant declination.
Furthermore, \cite{dlHOZ23} demonstrated that this correction can impact the recovery of foreground spectral parameters, such as $\beta_{S}$, in specific regions like the North Polar Spur when relying solely on QUIJOTE data. 

We then performed a robustness test comparing the same architecture trained with and without the FDEC-filtered maps, evaluating both cases on the same filtered dataset. This highlights the importance of incorporating this preprocessing step before estimating parameters from real data, which already has the filter applied. We found that, in general, only the lowest SNR sky regions and for the B-mode are more affected by not filtering the maps during training. The corresponding results are presented in Appendix \ref{sec:mode_loss}.

We also consider negligible Faraday Rotation effects in our simulations since, although there are a few contaminated pixels in the Galactic Plane, the maximum rotation angle at these regions at 11 GHz is of the order of 1 degree, with lower values at higher frequencies (\cite{RUB23}, \cite{dlHOZ23}).

We then generate three types of datasets: one for training the neural network, consisting on 10000 simulations, one for testing, formed by 1000 ones, and another for validation, constructed with 100 more simulations for each region. Each simulation centered on a given position (which is the same for all the 11, 13, 17, and 19 GHz QUIJOTE channels), is called Input Total patches in this work, plus the synchrotron map at 11 GHz, which is stored for each simulation as a label.

The simulations are thus constructed as follows: Firstly, we search the central position of each sky region. Then, the maps are cut into square patches of 256×256 pixels centered on those positions, with each pixel having the QUIJOTE angular resolution (approximately 6.9 arcmin). Next, point sources are injected into the patches, and a FWHM of 1$^{\circ}$ is applied. Finally, instrumental noise is added on each patch based on QUIJOTE MFI wide survey sensitivity (43.1, 38.2, 36.4 and 34.8 $\mu K$ per pixel on each channel \citep{GEN15}), which allows us to have different realizations for each simulation. Although simplistic, this noise model have been evaluated in previous works to be reliable to test this kind of methods on real data (\cite{CAS23}, \cite{CAS25}). However, we have tested for completeness in Appendix \ref{sec:noise_complexity} the performance of the neural network with different realizations of the QUIJOTE noise simulations, that have 1/$f$ term, several systematics and are anisotropic and have correlations between frequencies of the same horn \citep{RUB23}, finding that the correlated noise does not affect in general the estimated maps from the network.

For validation, both observations and simulations are cut into square patches using a central position of each region defined in \cite{FUS14}, having one patch per region. This allows to properly study the spectral variations on synchrotron emission across the sky. In the case of simulations, we cut 100 patches for each region. This is done by using 100 different maps with the same center point. In the case of the observations, we cut a single patch for each region using the same central point as for the simulations. 

An example of the patches used in this work at 11 GHz is shown in Figure \ref{fig:patches}, being the simulated sky at the top row, left column, the observations at the same sky location at the top row, center column, and the synchrotron model at the bottom row, left column.

The central position of these patches, reported in Table \ref{tab:table_regions} is taken to be approximately the center of each sky region defined in \cite{FUS14}, being the map publicly available\footnote{https://www.cosmoglobe.uio.no/products/cosmoglobe-dr1.html}. Each patch has then the same synchrotron physical properties and slighlty different than the other ones. Therefore, if the patch is smaller than the sky region, all the pixels could be used for analyzing the synchrotron parameters for such region. However, pixels from overlapping regions are masked to not consider them in the individual analysis of each region.

\section{Methodology}
\label{sec:methodology}

\begin{figure*}[ht]
    \centering
    \includegraphics[width=\linewidth]{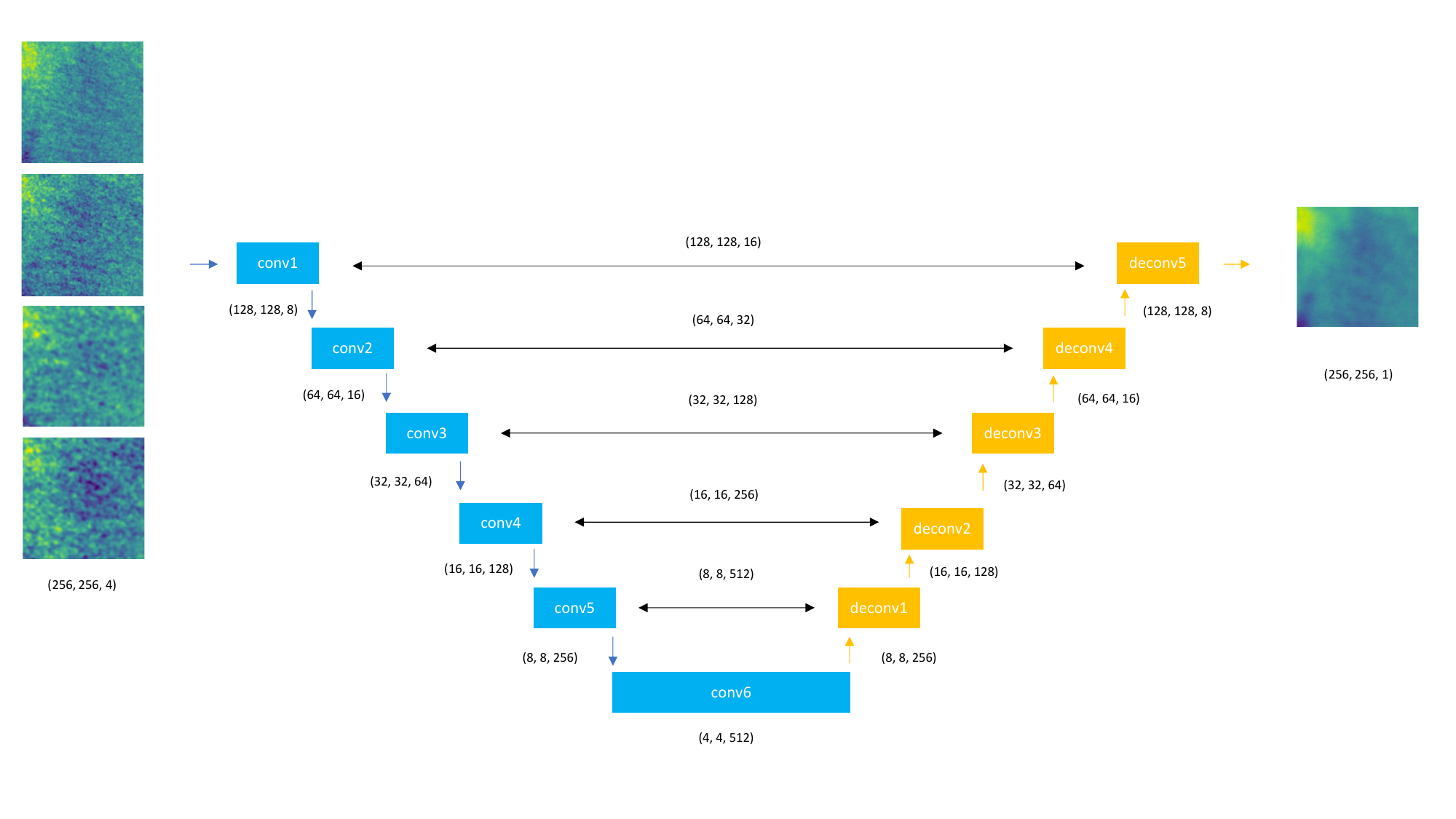}
    \caption{Scheme of the architecture of the neural network used in this work. We show for visualization purposes the kind of input and output patches the network reads and outputs, respectively.}
    \label{fig:network_architecture}
\end{figure*}

An artificial neural network is a machine learning framework composed of computational nodes, known as neurons, which are structured into multiple layers. Each neuron contains a set of weights that are adjusted during the training process. The input data propagate through the network. A loss function is located after the output map predicted by the neural network and it compares the label (the synchrotron s5 model at 11 GHz in the simulations) with respect to the output map, producing a numerical value called loss. This loss is minimized to compute a gradient that updates all the weights across the network using backpropagation. Once this update is complete, the process repeats for a new training cycle, known as an epoch. This iterative learning process then constitutes an optimization problem.

Neural networks are particularly effective for image-related tasks, since image data can be represented as two-dimensional arrays. A common application is image segmentation \citep{LON15}. In such cases, the learnable parameters are organized into matrices known as kernels. This kind of networks are then suitable for component separation processes, where the physical emissions present a high complexity and non-linearity. In this study, we adapt and extend the Cosmic Microwave Background Extraction Neural Network (CENN), initially introduced in \cite{CAS22b} to separate the CMB with respect to foregrounds in intensity maps, and then extended to polarization maps in \cite{CAS25}.

This network is a fully-convolutional one that builds upon the U-Net architecture \citep{RON15}. It is designed to reconstruct clean microwave maps from noisy background data. The theoretical foundations of such architectures for multi-frequency image processing are covered in \cite{GOO10}, while the physical ones are described in detail in Appendix \ref{sec:model}. While those references provide comprehensive explanations, we include here a concise summary of the network's structure and functionality. 

The network begins its training by ingesting the so-called Input Total patches, which contain sky maps at multiple frequencies. In this work, we aim to use QUIJOTE MFI data only, i.e. 11, 13, 17 and 19 GHz. This input data passes through six encoding blocks where convolutional filters are applied to extract features across various channels. An activation function introduces non-linearity to allow more complex patterns to be learned. After the encoding stage, the data passes through six decoding blocks, where deconvolutional operations reconstruct the signal. Each decoding block is linked to its corresponding encoder via skip connections, enabling the network to recover fine-grain details by comparing and integrating feature maps from both stages. The final layer outputs a patch representing the estimated map. This output is then compared with the label, which is the synchrotron s5 patch from the simulations at 11 GHz to compute the loss, which drives weight updates during training through gradient descent and backpropagation. For estimating the loss we have used the Mean Squared Error function and for estimating the gradient we have used the AdaGrad optimizer. The cycle then continues with the next epoch up to a total of 1000 epochs.

The full network architecture is depicted in Figure \ref{fig:network_architecture} and comprises six convolutional (encoder) blocks, each including convolution and pooling layers with kernel sizes of 9, 9, 7, 7, 5, and 3 pixels, and kernel counts of 8, 2, 4, 2, 2, and 2, respectively. These layers employ 8, 16, 64, 128, 256, and 512 filters and use a downsampling factor of 2. Padding type "Same" is applied throughout to preserve spatial dimensions. After that, six deconvolutional (decoder) blocks are connected and use similar deconvolution and pooling layers with kernel sizes of 3, 5, 7, 7, 9, and 9, and kernel counts of 2, 2, 2, 4, 2, and 8. The associated filter counts are 256, 128, 64, 16, 8, and 1. The same downsampling factor and padding strategy are used here, and all layers use the leaky ReLU activation function.

\section{Results}
\label{sec:results}

\subsection{Validation by simulations}
\label{sec:results_simulations}

\begin{figure*}[t]
\centering
\minipage{0.5\textwidth}
\includegraphics[width=\linewidth]{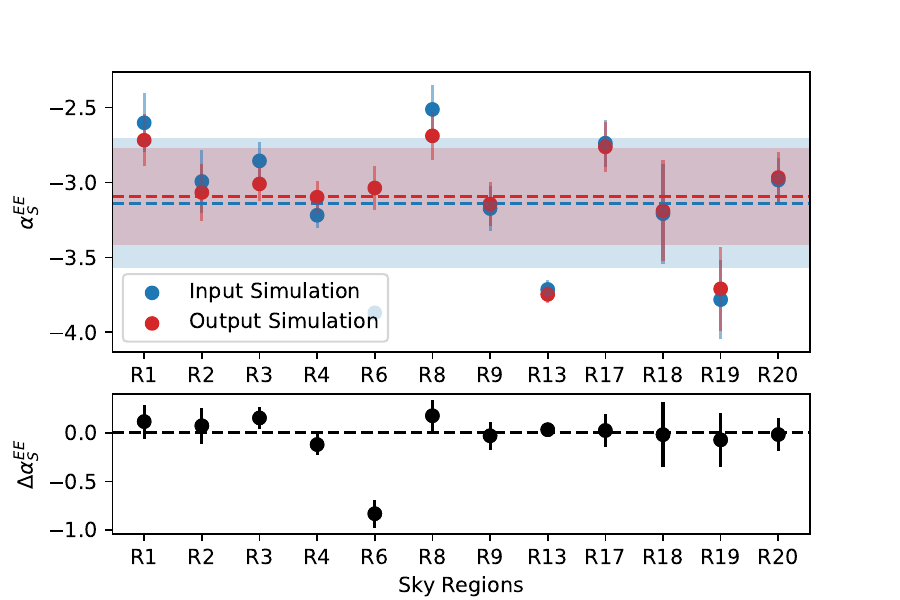}
\endminipage\hfill
\minipage{0.5\textwidth}%
  \includegraphics[width=\linewidth]{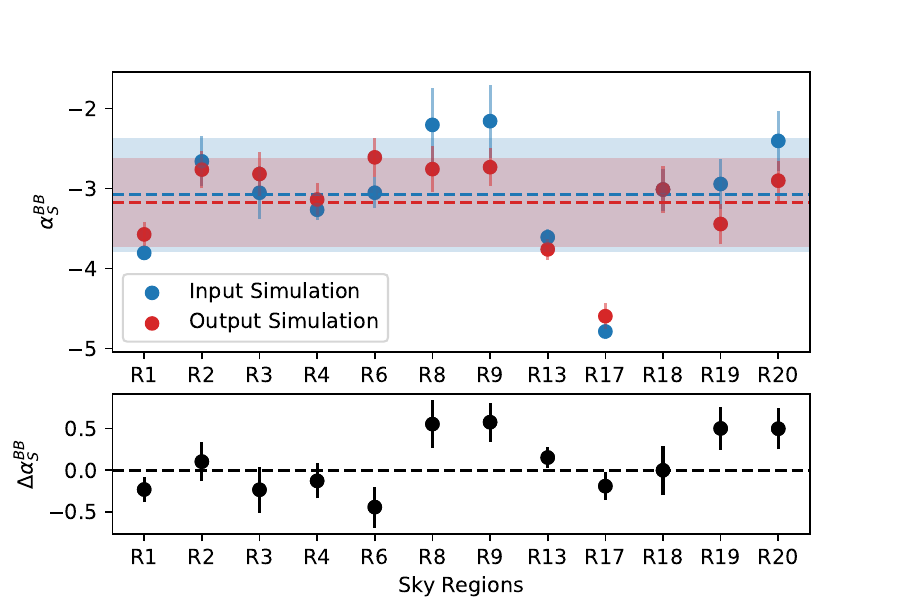}
\endminipage
\caption{Polarized synchrotron index $\alpha_{S}$ comparison at 11 GHz between the input s5 model (blue dots) and the outputs from the neural network (in red), after fitting both $EE$ (left panel) and $BB$ spectra (right panel) for each sky region, while bottom subpanel displays the absolute error between them. The average index is shown in both panels as a dashed blue and red lines, respectively. The average difference is represented as a dashed black line. The errorbars for each point show the 1$\sigma$ uncertainty for each sky region evaluated over 100 simulations. Shaded colored areas show the 1$\sigma$ uncertainties for each average value.}
\label{Fig:Simulations_SpectralIndex_ByRegion}
\end{figure*}

We consider for validation a dataset formed by 100 simulations for each sky region. For each region, we take the patches centered at the positions listed in Table \ref{tab:table_observations}. The network, trained individually for $Q$ and $U$, then outputs 100 synchrotron $Q$ and $U$ patches at 11 GHz for each region. An example of an output patch is shown in Figure \ref{fig:patches}, where the output from the network is shown at the middle row and the left column, and the difference with respect input, i.e. the synchrotron s5 model, is displayed at the bottom row, middle column. 

We then estimate the average $EE$ and $BB$ power spectra for each 100 region patches using NaMaster \citep{NAMASTER}, using the apodization, mask and binning described in the main code\footnote{https://github.com/LSSTDESC/NaMaster}, and we fit the data by assuming a power law model as described in \eqref{eq:power_law}, and following the parametrization defined in \eqref{eq:power_law_parametrization}, for multipoles in the range $50 \lesssim l \lesssim 300$. The fitting is performed by using the curve fit method from Scipy \citep{scipy}. We then obtain estimates of the synchrotron amplitude $A_{S}$ and the index $\alpha_{S}$ for both $EE$ and $BB$, and then we can estimate the ratio of the amplitudes, $r = {A_{S}^{BB}}/{A_{S}^{EE}}$. 

Figure \ref{Fig:Simulations_SpectralIndex_ByRegion} presents the comparison between the synchrotron $\alpha_{S}$ index in the input s5 simulations (in blue) with respect the estimates from the neural network (in red) at 11 GHz in the simulated QUIJOTE sky, for $EE$ and $BB$ in the left and right panels, respectively, and for each sky region. The index absolute error is shown in the bottom subpanel in black. The error bars for each estimation represent the 1$\sigma$ standard deviation for the 100 simulations. The dashed horizontal lines represent the average estimates and the shaded areas display the 1$\sigma$ standard deviation for these average values.

Overall, the neural network captures relatively well the spatial behavior of the index in $EE$, as well as the average one, although some biases are found in $BB$. Particularly, there is a strong bias for $EE$ in Region 6, and several ones in $BB$, especially in Regions 8, 9, 19 and 20, suggesting localized reconstruction challenges, probably due to low signal-to-noise ratio (SNR, Regions 6, 8 and 9) and/or complex spatial structure (Regions 18 and 20). 

In summary, the neural network demonstrates overall accuracy in recovering the polarized synchrotron index $\alpha_{S}$, with better performance in the $E$-mode component. The localized biases identified in certain $B$-mode regions deserve further investigation and may point to Galactic areas where model refinement could improve generalization and robustness. Future trainings should use data at frequencies where synchrotron SNR is higher in some regions.

\begin{figure*}[t]
\centering
\minipage{0.5\textwidth}
\includegraphics[width=\linewidth]{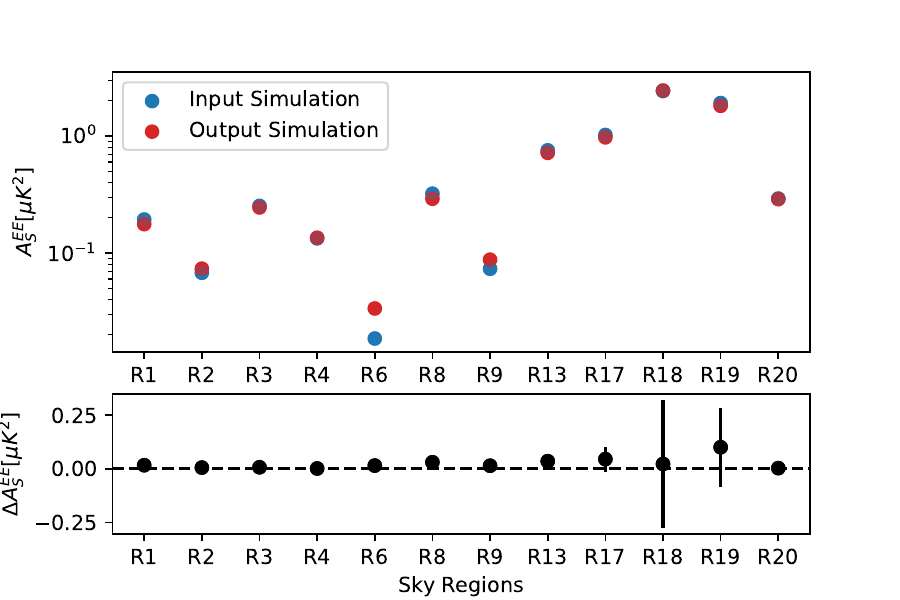}
\endminipage\hfill
\minipage{0.5\textwidth}%
  \includegraphics[width=\linewidth]{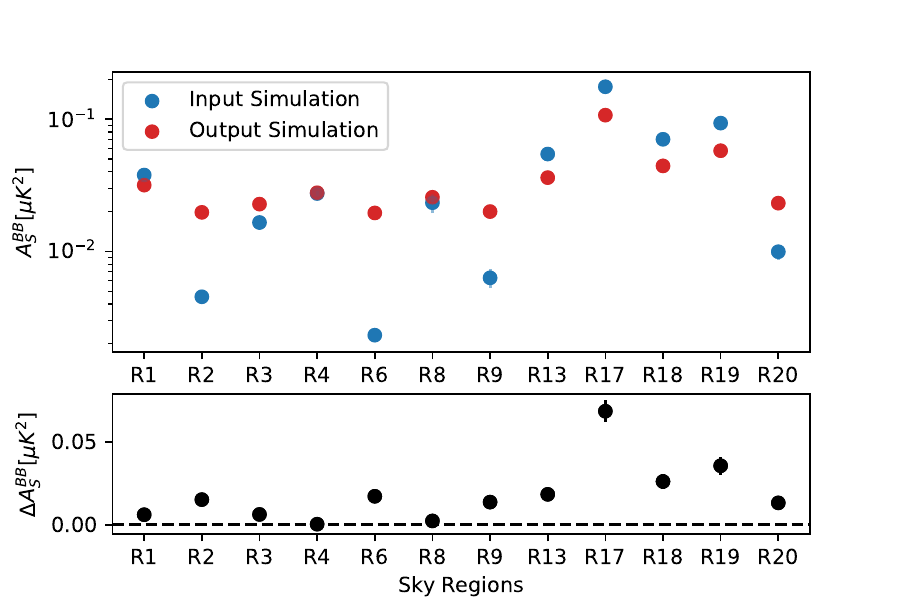}
\endminipage
\caption{Polarized synchrotron amplitude $A_{S}$ comparison at 11 GHz in logarithmic scale between the input s5 model (blue dots) and the outputs from the neural network (in red), after fitting both $EE$ (left panel) and $BB$ spectra (right panel) for each sky region, while bottom subpanel displays in linear scale the absolute error between them. The average difference is represented as a dashed black line. The errorbars for each point show the 1$\sigma$ uncertainty for each sky region evaluated over 100 simulations. Shaded colored area show the 1$\sigma$ uncertainty for the absolute error average value.}
\label{Fig:Simulations_Amplitude_ByRegion}
\end{figure*}

Figure \ref{Fig:Simulations_Amplitude_ByRegion} presents the comparison between the synchrotron $A_{S}$ amplitude in the input s5 simulations (in blue) with respect the estimates from the neural network (in red) at 11 GHz in the simulated QUIJOTE sky, for $EE$ and $BB$ in the left and right panels, respectively, and for each sky region. The index absolute error is shown in the bottom subpanel in black. The error bars for each estimation represent the 1$\sigma$ standard deviation for the 100 simulations. The dashed horizontal lines represent the average estimates and the shaded areas display the 1$\sigma$ standard deviation for these average values. 

As shown, in both polarization modes, the network outputs follow the overall trend of the last section, i.e. for $EE$, the estimated amplitudes are in general similar than the input ones, mainly due to the high SNR at 11 GHz in most of the regions for this mode. However, as in the last section, Region 6 remains the one with the higher discrepancies with respect to the input one, because the low SNR of this sky region. For $BB$, there are more estimates that present strong biases, especially regions with low SNR as R2, R6 and R9.

In summary, the neural network exhibits also for this quantity relatively robust performance in recovering the amplitude of polarized synchrotron emission in both $E$ and $B$-mode components, but the SNR of the simulated instrument as well as the sky signal in some sky regions make it difficult some estimations, especially for $BB$. As for the synchrotron index estimation, the accuracy of this estimation would likely improve with the addition of training data with higher sensitivity. This will be evaluated in a future work.

\begin{figure}[ht]
\centering
\includegraphics[width=\linewidth]{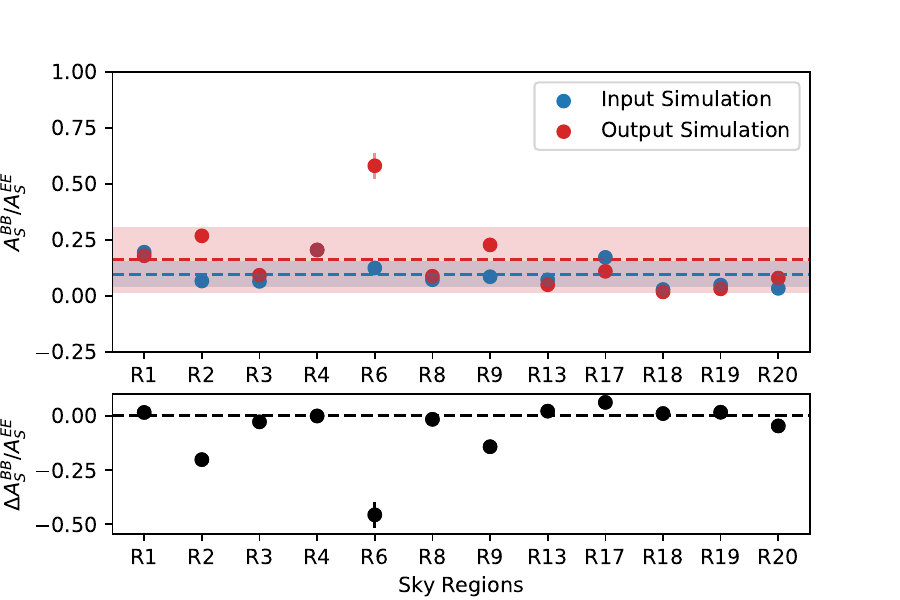}
\caption{Polarized synchrotron ratio comparison between $B$ and $E$ amplitudes at 11 GHz between the input s5 model (blue dots) and the outputs from the neural network (in red), after fitting both spectra for each sky region. The bottom subpanel displays the absolute error between them. The average ratio is shown as a dashed blue and red lines, respectively. Shaded colored areas show the 1$\sigma$ respective uncertainties. The average difference is represented as a dashed black line. The errorbars for each point show the 1$\sigma$ uncertainty for each sky region propagated from $A_{S}^{BB}$ and $A_{S}^{EE}$ uncertainties using equation \ref{eq:error_propagation} evaluated over 100 simulations. Shaded colored area show the 1$\sigma$ uncertainty for the absolute error average value.}
\label{Fig:Simulations_r_ByRegion}
\end{figure}

Finally, Figure \ref{Fig:Simulations_r_ByRegion} shows the comparison between the synchrotron $BB$ and $EE$ amplitude ratio in the input simulations (in blue) with respect to the ratio from the outputs of the neural network (in red). Their absolute difference is shown in the bottom subpanel in black for each sky region. The errorbars for each estimation show the 1$\sigma$ uncertainty by propagating the errors from the 100 simulations using
\begin{equation}
\delta r = r \cdot \sqrt{\left(\frac{\delta A_{BB}}{A_{BB}}\right)^2 + \left(\frac{\delta A_{EE}}{A_{EE}}\right)^2},
\label{eq:error_propagation}
\end{equation}
where $\delta r$ is the uncertainty in $r$, $\delta A_{XX}$ is the uncertainty in the amplitudes and $A_{XX}$ are the amplitudes.

The dashed horizontal line represents the mean ${A_{S}^{BB}}/{A_{S}^{EE}}$ ratios, averaged from all the regions (and assumed to be the one at all sky) and then the shaded areas presents the 1$\sigma$ uncertainties of these values.

Overall, the neural network seems to successfully reconstruct the ratio with good agreement to the input simulations in most sky regions. The predicted values closely follow the input distribution, with the majority of the points lying within the uncertainty range. The mean of the output values remains consistent with that of the input, suggesting that the model captures the global behavior of the synchrotron polarization ratio, although some regions such R6 display an unconsistent value between input and output. These discrepancies may stem from the intrinsic variability and low signal-to-noise ratio of the synchrotron $B$-mode component, which is more challenging to accurately model.

In summary, the network demonstrates a strong ability to reconstruct the synchrotron polarization amplitude ratio across different sky regions, although bias, that should carefully consider appear in low SNR regions.

\subsection{Application to real QUIJOTE MFI data}
\label{sec:results_observations}

Once we deduce that the results obtained by the neural network are generally consistent with the input simulations, we decide to test its performance on the QUIJOTE MFI real observations, publicly available. An example of the input and output patches of the QUIJOTE MFI observations for the North Polar Spur (Region 13) at 11 GHz is shown in Figure \ref{fig:patches}, being the input sky represented at the top row, middle column and the output from the network shown at the middle row and column. Also the difference with respect the synchrotron s5 model is plotted at the bottom row, right column. Moreover, we show in the middle row and right column the difference between the output from the network and the simulations at the same point in the sky. For completeness, the same but for the input sky is displayed in the top row, right column.

The main insights extracted from this particular example is that it seems to be a difference at the middle scales between the synchrotron s5 model and the QUIJOTE MFI observations, which are probably synchrotron structures not properly modeled. Moreover, once the network is trained, after comparing the difference between its outputs from the simulations and the observations, it seems that it is capable of generalize to different data since at the bottom row and middle column of Figure \ref{fig:patches}, the network is mainly recovering the synchrotron emission with some residual instrumental noise, while for the observations it is recovering such synchrotron structures that were not in the training nor validating data. 

\begin{figure*}[t]
\centering
\minipage{0.5\textwidth}
\includegraphics[width=\linewidth]{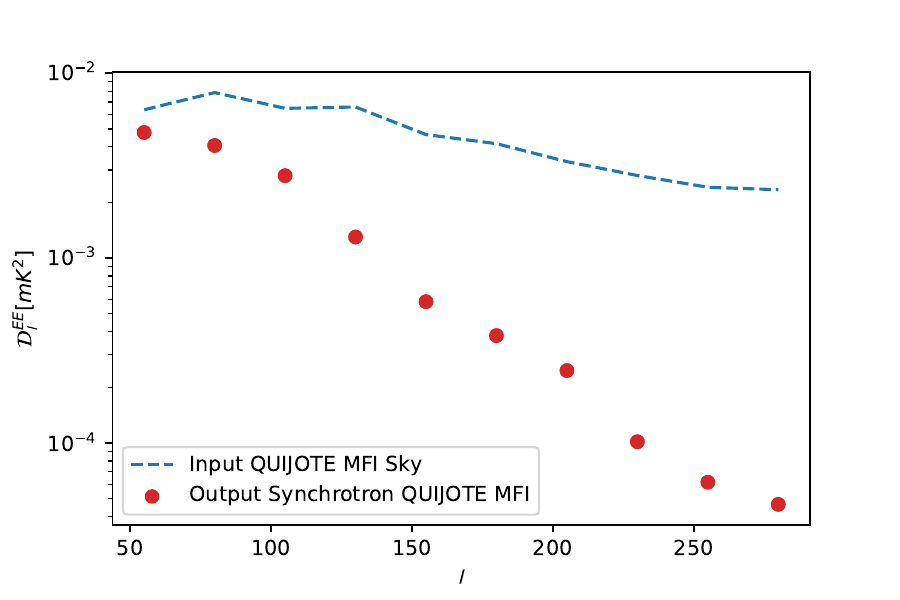}
\endminipage\hfill
\minipage{0.5\textwidth}%
  \includegraphics[width=\linewidth]{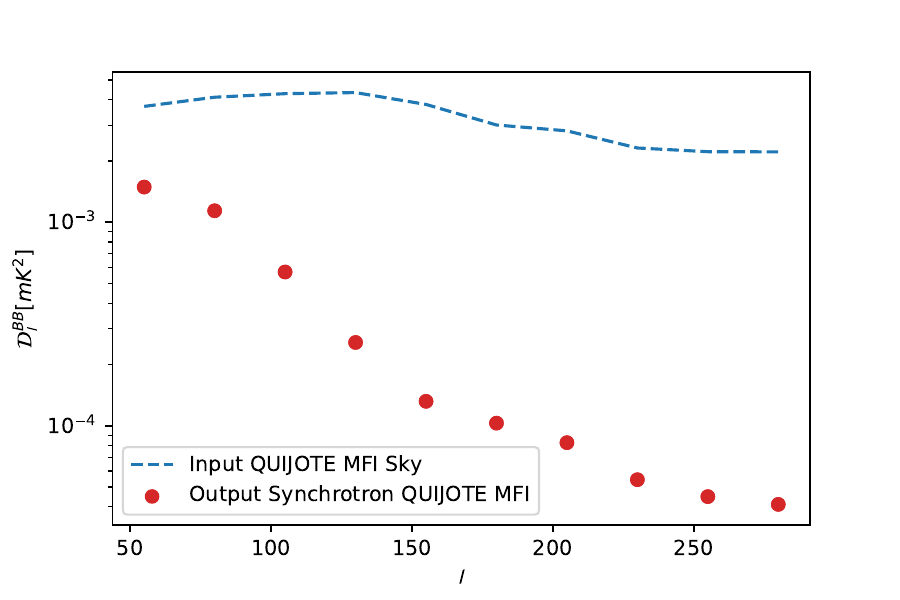}
\endminipage
\caption{Power spectra comparison at 11 GHz between the output synchrotron averaged over all the 12 sky regions (in red dots) with respect to the power spectra of the QUIJOTE MFI sky (in blue dashed line) for $EE$ (left panel) and $BB$ (right panel).}
\label{Fig:Power_Spectrum}
\end{figure*}

We then estimate both $EE$ and $BB$ spectra over each one of the 12 sky patches to estimate the synchrotron parameters at 11 GHz in the same way as for simulations. We plot their average value from all regions in Figure \ref{Fig:Power_Spectrum} at the left and right panels, respectively, in comparison with respect to the average power spectra of the QUIJOTE MFI sky (i.e. sky signal plus instrumental noise), which is some orders of magnitude higher at middle scales due mainly to instrumental noise. As shown, the spectra presents in both cases a power-law shape, which is more clear for $EE$ due to its higher SNR with respect to $BB$. Therefore, no curvature signals seem to be visually present in this work and the addition of more data from higher frequencies (WMAP and \textit{Planck}) and lower ones (CBASS) becomes mandatory for such analysis.

\begin{figure*}[t]
\centering
\minipage{0.5\textwidth}
\includegraphics[width=\linewidth]{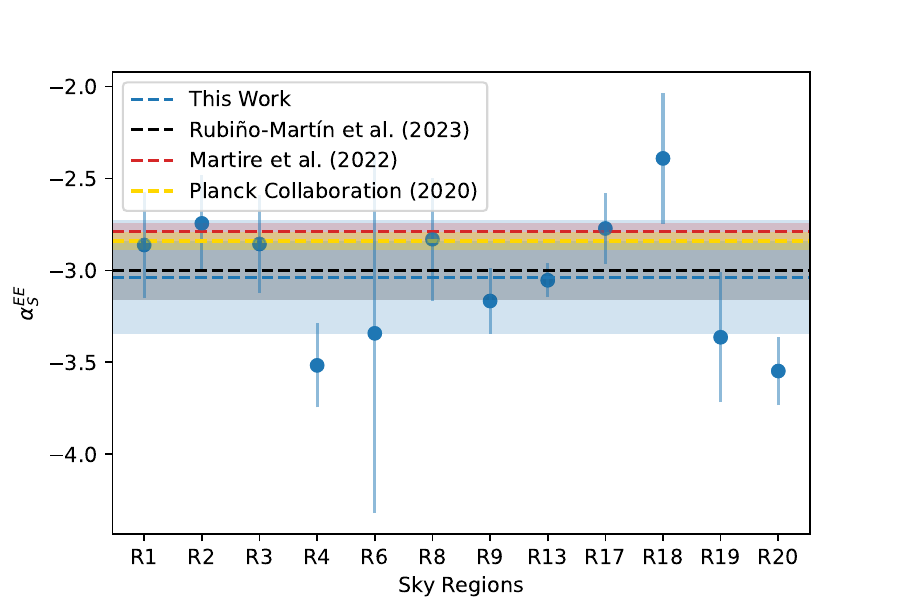}
\endminipage\hfill
\minipage{0.5\textwidth}%
  \includegraphics[width=\linewidth]{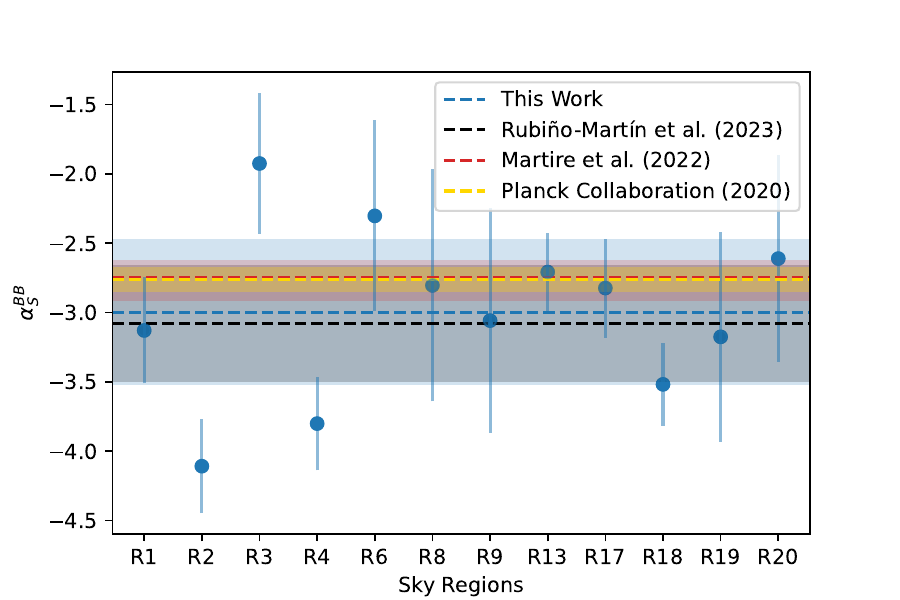}
\endminipage
\caption{Polarized synchrotron index $\alpha_{S}$ estimates (blue dots) in the QUIJOTE MFI observations at 11 GHz by fitting $EE$ (left panel) and $BB$ spectra (right panel) for each sky region. The average index is shown in both panels as a dashed blue line. The errorbars for each point represents both the statistical and systematical uncertainties for each sky region. As a reference, we show the comparison with \cite{RUB23} (black dashed line), \cite{MAR22} (red dashed line) and \cite{PLA_18_IV} (yellow dashed line) at similar sky mask coverages than in this work. Shaded colored areas display the respective 1$\sigma$ uncertainties for every average estimation.}
\label{Fig:SpectralIndex_ByRegion}
\end{figure*}

\begin{figure*}[t]
\centering
\minipage{0.5\textwidth}
\includegraphics[width=\linewidth]{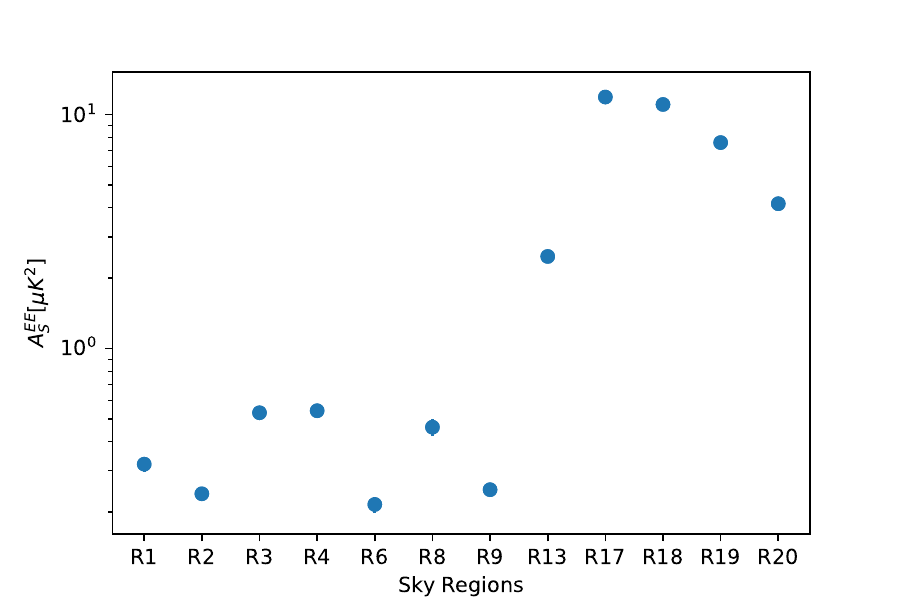}
\endminipage\hfill
\minipage{0.5\textwidth}%
  \includegraphics[width=\linewidth]{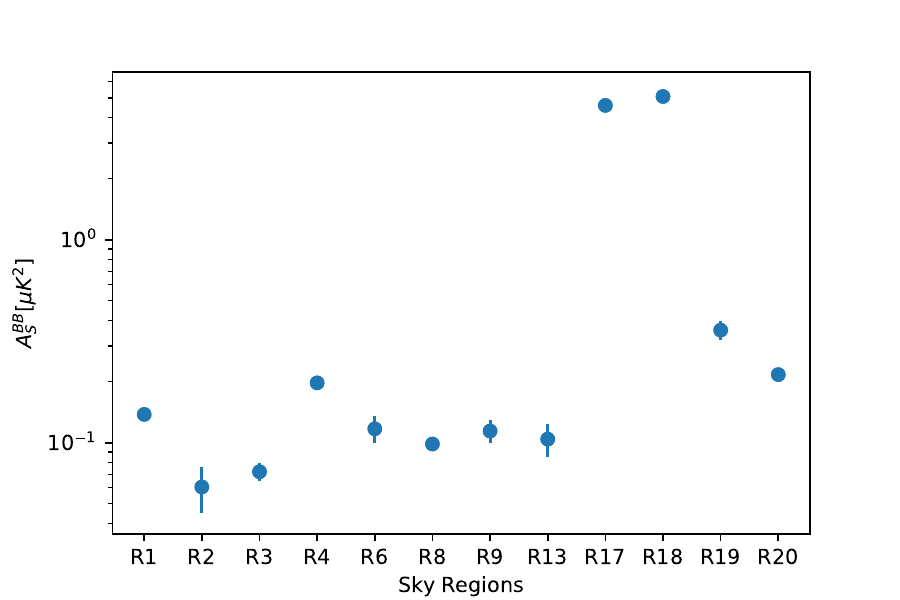}
\endminipage
\caption{Polarized synchrotron amplitude $A_{S}$ estimates in logarithmic scale in the QUIJOTE MFI observations at 11 GHz after fitting $EE$ (left panel) and $BB$ spectra (right panel) for each sky region. The errorbars for each point represents both the statistical and systematical uncertainties for each sky region.}
\label{Fig:Amplitude_ByRegion}
\end{figure*}

\begin{figure}[ht]
\centering
\includegraphics[width=\linewidth]{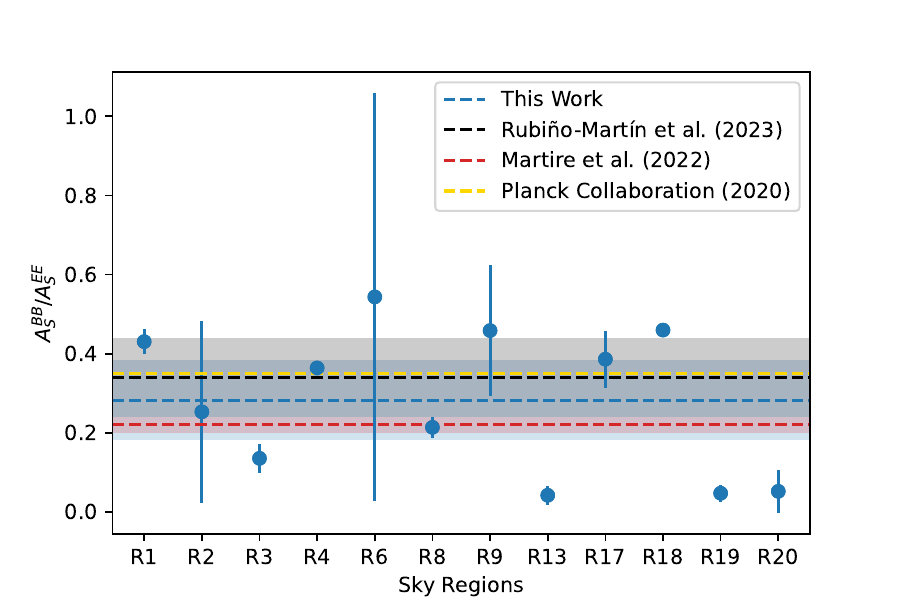}
\caption{Polarized synchrotron estimates for the ratio between $A_{S}^{BB}$ and $A_{S}^{EE}$ amplitudes (blue dots) in the QUIJOTE MFI observations at 11 GHz for each sky region. The errorbars for each point show both statistical and systematical uncertainties for each sky region propagated from $A_{S}^{BB}$ and $A_{S}^{EE}$ uncertainties using equation \ref{eq:error_propagation}. The average ratio is displayed as a dashed blue line. As a reference, we show the comparison with \cite{PLA_18_IV} (yellow dashed line), \cite{MAR22} (red dashed line) and \cite{RUB23} (black dashed line) at similar sky mask coverages than in this work. Shaded colored areas display the respective 1$\sigma$ uncertainties for every average estimation.}
\label{Fig:r_ByRegion}
\end{figure}

Based on the average spectra, we then fit it individually for each sky region by assuming a power-law behavior by following equation \ref{eq:power_law}. Results are summarize in Table \ref{tab:table_observations}, and represented in Figures \ref{Fig:SpectralIndex_ByRegion}, \ref{Fig:Amplitude_ByRegion} and \ref{Fig:r_ByRegion}. Appendix \ref{sec:full_sky_maps} shows the results represented in all sky maps divided into regions for visualization purposes.

\begin{table*}[ht]
    \caption{Summary of our synchrotron estimates in the QUIJOTE MFI observations at 11 GHz for each sky region. From left to right, columns show: the sky region analyzed, for which patch central position is displayed in Table \ref{tab:table_regions}, the estimated amplitudes $A_{S}^{EE}$ and $A_{S}^{BB}$, the indeces $\alpha_{S}^{EE}$ and $\alpha_{S}^{BB}$, and finally the ratio between the amplitudes $A_{S}^{BB}/A_{S}^{EE}$. The data show on each row both the statistical and systematical uncertainties, respectively, also averaged for the Galactic Plane regions (GP), high-latitude regions (HLR) and all regions, as defined in Table \ref{tab:table_regions}.}
    \label{tab:table_observations}
    \centering
\begin{tabular}{cccccc} 
\hline
Region  & $A_{S}^{EE}$ & $A_{S}^{BB}$ & $\alpha_{S}^{EE}$ & $\alpha_{S}^{BB}$ & $A_{S}^{BB}/A_{S}^{EE}$ \\
\hline
1  & 0.32 $\pm$ 0.03 & 0.14 $\pm$ 0.02 & -2.86 $\pm$ 0.21 & -3.13 $\pm$ 0.27 & 0.44 $\pm$ 0.02 \\
2  & 0.24 $\pm$ 0.02 & 0.06 $\pm$ 0.02 & -2.7 $\pm$ 0.21 & -4.1 $\pm$ 0.22 & 0.25 $\pm$ 0.20 \\
3  & 0.53 $\pm$ 0.02 & 0.07 $\pm$ 0.02 & -2.86 $\pm$ 0.19 & -1.9 $\pm$ 0.36 & 0.14 $\pm$ 0.03 \\
4  & 0.54 $\pm$ 0.02 & 0.20 $\pm$ 0.01 & -3.5 $\pm$ 0.14 & -3.8 $\pm$ 0.22 & 0.36 $\pm$ 0.02 \\
6  & 0.21 $\pm$ 0.02 & 0.12 $\pm$ 0.02 & -3.3 $\pm$ 0.81 & -2.3 $\pm$ 0.50 & 0.50 $\pm$ 0.46 \\
8  & 0.46 $\pm$ 0.04 & 0.10 $\pm$ 0.01 & -2.83 $\pm$ 0.24 & -2.8 $\pm$ 0.58 & 0.21 $\pm$ 0.02 \\
9  & 0.25 $\pm$ 0.02 & 0.10 $\pm$ 0.01 & -3.17 $\pm$ 0.15 & -3.1 $\pm$ 0.63 & 0.46 $\pm$ 0.14 \\
13 & 2.50 $\pm$ 0.11 & 0.10 $\pm$ 0.02 & -3.00 $\pm$ 0.06 & -2.71 $\pm$ 0.18 & 0.04 $\pm$ 0.03 \\
17 & 11.9 $\pm$ 0.50 & 4.60 $\pm$ 0.20 & -2.77 $\pm$ 0.17 & -2.82 $\pm$ 0.25 & 0.39 $\pm$ 0.06 \\
18 & 11.1 $\pm$ 0.53 & 5.10 $\pm$ 0.21 & -2.39 $\pm$ 0.34 & -3.50 $\pm$ 0.30 & 0.46 $\pm$ 0.01 \\
19 & 7.6 $\pm$ 0.37 & 0.36 $\pm$ 0.04 & -3.36 $\pm$ 0.29 & -3.18 $\pm$ 0.56 & 0.05 $\pm$ 0.02 \\
20 & 4.17 $\pm$ 0.17 & 0.21 $\pm$ 0.02 & -3.55 $\pm$ 0.17 & -2.61 $\pm$ 0.56 & 0.05 $\pm$ 0.05 \\
\hline
GP  & 7.60 $\pm$ 0.41 & 1.88 $\pm$ 0.11 & -3.10 $\pm$ 0.30 & -3.10 $\pm$ 0.28 & 0.18 $\pm$ 0.01 \\
HLR & 0.63 $\pm$ 0.08 & 0.11 $\pm$ 0.01 & -3.05 $\pm$ 0.16 & -2.98 $\pm$ 0.23 & 0.30 $\pm$ 0.10 \\
All & 3.31 $\pm$ 0.17 & 0.93 $\pm$ 0.04 & -3.04 $\pm$ 0.18 & -3.00 $\pm$ 0.26 & 0.28 $\pm$ 0.06 \\
\hline
\end{tabular}
\end{table*}

In particular, Figure \ref{Fig:SpectralIndex_ByRegion} presents our estimations for the synchrotron $\alpha_{S}$ index in the QUIJOTE real observations at 11 GHz after fitting both $EE$ (left panel) and $BB$-mode spectra (right panel) with the same conditions than the ones described in Section \ref{sec:results_simulations} with the simulations, and for each sky region. The errorbars for each estimation show the statistical+systematical uncertainty, where the statistical is the one from the 100 simulations on each regions presented in the last section and the systematical is the bias between input s5 and output also showed in the last section for the same simulations. We have also assessed the impact of adopting alternative synchrotron models, such as the s7 model, on the derived uncertainties. However, based on the results presented in Figures \ref{Fig:Index_s7}, \ref{Fig:Amplitude_s7}, and \ref{Fig:r_s7}, which are consistent with those obtained in this Section, we opt to retain only the s5-based analysis for the purpose of defining the uncertainties.

Moreover, the dashed horizontal line represents the mean estimated synchrotron index, averaged from all the regions (and assumed to be the one at all sky), in comparison with respect to previous similar analyses (\cite{RUB23} (black dashed line), \cite{MAR22} (red dashed line) and \cite{PLA_18_IV} (yellow dashed line)). Shaded areas show the 1$\sigma$ uncertainty for each average value, being in our case the sum of statistical and systematical uncertainty value for this average. We did not compare with \cite{KRA18} because of their different sky coverage with respect to ours, although as described in Section \ref{sec:introduction}, we found compatible average results.

The estimated average values of the synchrotron indeces are $\alpha_{S}^{EE} = -3.04 \pm 0.18$ and $\alpha_{S}^{BB} = -3.00 \pm 0.26$, consistent with all the previous analyses, especially with the one using the same data \citep{RUB23}, which found $\alpha_{S}^{EE} = -3.00 \pm 0.16$ and $\alpha_{S}^{BB} = -3.08 \pm 0.42$ for a Galactic mask of $|b|>5^{\circ}$.

Furthermore, the index shows a clear spatial variability along the sky with some of the regions showing a flatter index while others presenting a steeper one. There are also regions showing high discrepancies between $\alpha_{S}^{EE}$ and $\alpha_{S}^{BB}$, some of them indeed inside the errorbars, thus we can infer that some biases from training the network with a simplistic synchrotron model and/or noise, and other ones outside the errorbars (R2, R3, R18), that we relate to either structures not reconstructed in the $B$-mode (R18) and/or low SNR regions (R2 and R3). 

Moreover, Figure \ref{Fig:Amplitude_ByRegion} displays our estimates for the synchrotron $A_{S}$ amplitude in the QUIJOTE MFI real observations at 11 GHz, after fitting both $EE$ (left panel) and $BB$-mode spectra (right panel), and for each sky region. The errorbars for each estimation show the statistical+systematical uncertainty, calculated as in the previous analysis, although for these estimates, we also consider the relative 4\% calibration uncertainty described in \cite{RUB23} for polarization, as a quadratic sum of the statistical and systematic uncertainties. We have also included for region 13 (i.e. the North Polar Spur) the error due to not filtering the maps with the FDEC of the training dataset, as explained in Appendix \ref{sec:mode_loss}, also as a quadratic sum of the other uncertainties.

We estimate average amplitudes considering all the sky regions of $A_{S}^{EE} = 3.31 \pm 0.17 \mu K^{2}$ and $A_{S}^{BB} = 0.93 \pm 0.04 \mu K^{2}$. In the Galactic Plane, we find $A_{S}^{EE} = 7.60 \pm 0.41 \mu K^{2}$ and $A_{S}^{BB} = 1.88 \pm 0.11 \mu K^{2}$ and in the outer regions $A_{S}^{EE} = 0.63 \pm 0.08 \mu K^{2}$ and $A_{S}^{BB} = 0.11 \pm 0.01 \mu K^{2}$. Our estimates, indeed, are also close to the \cite{RUB23} ones if we compare their Galactic cuts with ours (an average $A_{S}^{EE} = 1.52 \pm 0.15 \mu K^{2}$ and $A_{S}^{BB} = 0.52 \pm 0.15 \mu K^{2}$ for $|b| > 5^{\circ}$, and $A_{S}^{EE} = 0.81 \pm 0.19 \mu K^{2}$ and $A_{S}^{BB} = 0.18 \pm 0.13 \mu K^{2}$ for $|b| > 20^{\circ}$).

Although the estimates follow the expected behavior, i.e. regions with low SNR would present the lower amplitudes and the same for the high SNR regions, it  seems to be an overestimation of the Galactic region amplitudes since R13 appears to be lower than theirs, and should be as known the region with the highest SNR. This overestimation is likely due to residual instrumental noise in the maps, and therefore it is expected to have more accurate estimates either joining data from other surveys and/or with future observations with higher sensitivity as QUIJOTE MFI2 ones.

Finally, Figure \ref{Fig:r_ByRegion} shows our estimates for the ratio between $B$ and $E$ amplitudes in the QUIJOTE MFI real observations at 11 GHz. The errorbars for each estimation show the statistical+systematical uncertainty. The statistical one have been estimated by propagating the errors from the ratio ${A_{S}^{BB}}/{A_{S}^{EE}}$ in the simulations. The systematic one is given by the bias between input and output ratio in the simulations.

The dashed horizontal line represents the average ${A_{S}^{BB}}/{A_{S}^{EE}}$ ratio from all the regions (and assumed to be the one at all sky). As a comparison, we show the results from previous analyses, \cite{RUB23} (in a black dashed line), \cite{MAR22} (in a red dashed line) and \cite{PLA_18_IV} (in a yellow dashed line). Shaded areas show the 1$\sigma$ uncertainty for each average value.

We find an average ratio of $0.28 \pm 0.06$, which is closer to the \cite{RUB23} one ($0.34 \pm 0.10$ for $|b| > 5^{\circ}$). Also, we can compare with other analyses such \cite{PLA_18_IV} ($\approx0.39$), \cite{MAR22} ($\approx0.22$), although they used data at higher frequencies. As shown, our value is also consistent with them but they had lower uncertainty, likely due to the combination of \textit{Planck} and WMAP data instead of using QUIJOTE data only.

As in the previous sections, low SNR regions such R2, R6 and R9 present the highest uncertainties, as expected, due to the error propagation from ${A_{S}^{BB}}$ and ${A_{S}^{EE}}$.

\section{Discussion and conclusions}
\label{sec:conclusions}

In this work, we aimed to analyze the scale-dependent spatial variation properties of the polarized synchrotron emission in the QUIJOTE MFI data by fitting both $EE$ and $BB$ spectra on output maps from a novel component separation technique called CENN, presented firstly in \cite{CAS22b} with intensity maps, and then extended to polarization ones in \cite{CAS24}.

As said before, our analysis aims to extract reliable estimates of synchrotron parameters, such as the synchrotron index $\alpha_{S}$ and the amplitude for both $E$ and $B$-modes, as so as the ratio between $B$ and $E$ amplitudes. The spatial variability will be studied by considering several physically distinct sky regions, defined in \cite{FUS14}. Although using QUIJOTE data alone instead of making joint analyses with \textit{Planck} and WMAP could led to huge uncertainties, we wanted to study the performance of the neural network dealing with this kind of data to give preliminary estimations of the synchrotron properties in the northern hemisphere, before training other neural network with more complete data.

To train and validate the neural network, we have constructed a set of realistic simulations that incorporate all the elements forming the microwave sky at the frequencies covered by the QUIJOTE MFI instrument, i.e. synchrotron, dust, AME, the CMB and radio point sources, and also we added instrumental noise following the sensitivity specifications of the QUIJOTE MFI instrument, but without taking into account systematic effects, although we know that this could be a simplification with respect the real data in polarization, but we did not want to make a complex training in this first analysis. Synchrotron emission was modeled with the s5 PySM configuration, which assumes a spatial variability in the spectral index between the Galactic plane and the extragalactic regions, and also incorporates small scale features. We have not used a curvature model in the training dataset because in this work we are assuming a power-law behavior but we performed a test with the trained network and data with a synchrotron curvature model, described in Appendix \ref{sec:curvature_models}.

The simulations were cut into square patches of 256$\times$256 pixels at the QUIJOTE's angular resolution (approximately 414 arcseconds per pixel), covering the four MFI frequencies of 11, 13, 17, and 19 GHz. Each sky region was populated with 100 independent simulations, allowing for robust uncertainty estimation. For the observational analysis, we employed the publicly available QUIJOTE MFI maps, extracting one patch per region, centered on the same positions used in the simulations.

After being trained with patches at random positions in the sky, the neural network demonstrated relatively accurate performance when recovering the synchrotron signal from simulated sky patches, in particular for the synchrotron index for both polarization modes, with average absolute errors below 0.5 in most of the sky regions, and generally better performance for $E$ compared to the $B$-mode. 

Once validated using realistic simulations, the network was subsequently applied to real QUIJOTE-MFI data. Despite being trained exclusively on synthetic maps, the network appears to reliably extract meaningful synchrotron structures from actual sky observations. Also, mid-scale features not present in the training data model. This will be further investigated in future works. 

The average values from all the regions are in general consistent with previous analyses. We found $\alpha_{S}^{E} = -3.04 \pm 0.18$ and $\alpha_{S}^{B} = -3.00 \pm 0.26$, amplitudes of $A_{S}^{E} = 3.31 \pm 0.17 \mu K^{2}$ and $A_{S}^{B} = 0.93 \pm 0.04 \mu K^{2}$, and a ratio between $B$ and $E$ amplitudes of $A_{S}^{B}/A_{S}^{E} = 0.28 \pm 0.06$, considering systematic+statistical uncertainties in all the cases.

We also found notable spatial variations in the synchrotron properties output by the network, as previous analyses found for other microwave experiments and frequencies. In particular, we estimate that, at high Galactic latitudes (R1-R13), synchrotron generally exhibited flatter indices, with average values of $\alpha_{S}^{E} = - 3.05 \pm 0.16$ and $\alpha_{S}^{B} = - 2.98 \pm 0.23$, lower amplitudes ($A_{S}^{E} = 0.63 \pm 0.08 \mu K^{2}$ and $A_{S}^{B} = 0.11 \pm 0.01 \mu K^{2}$), and higher $B$ to $E$ amplitude ratios ($A_{S}^{B}/A_{S}^{E} = 0.3 \pm 0.1$). On the other hand, Galactic plane regions (R17–R20) showed steeper indices , with average values of $\alpha_{S}^{E} = - 3.1 \pm 0.3$ and $\alpha_{S}^{B} = - 3.10 \pm 0.28$, higher amplitudes ($A_{S}^{E} = 7.60 \pm 0.41 \mu K^{2}$ and $A_{S}^{B} = 1.88 \pm 0.11 \mu K^{2}$), and lower $B$ to $E$ amplitude ratios ($A_{S}^{B}/A_{S}^{E} = 0.18 \pm 0.01$), indicating strong $E$-mode dominance, as expected.

In conclusion, as we know, the physical behavior of the Interstellar Medium is extremely complex in polarization at these frequencies. Then, overall, this work is a huge step on the validation of these methods for their use in cosmic microwave background present and future data. We have presented a methodology which can be used for foreground cleaning and separation, in order to construct foreground templates to help to optimize component separation methods before future observations, as so as, with the properly data and simulations, it could be used also as a complementary component separation method itself with respect parametric and semi-blinds ones in order to give robust estimates with different kind of methodologies.

Future work, as said before, will focus on extending this approach by incorporating additional data from WMAP, \textit{Planck} LFI, and/or upcoming survey data as CBASS, as so as improving the foreground and instrumental noise modeling. This will enhance the robustness of the network across a broader frequency range, improve performance in low signal regions, and help to mitigate systematic uncertainties.

\begin{acknowledgements}
We warmly thank the anonymous referee for the constructive comments that improved the clarity of this work. JMC, LB and JGN acknowledge the CNS2022-135748 / MCIN/AEI/10.13039/501100011033 project. JMC, LB, JGN, DC, RFF and JAC acknowledge the PID2021-125630NB-I00 / MCIN/AEI/10.13039/501100011033 project. JARM and RGS acknowledge the PID2023-151567NB-I00 project. JMC, JARB, RTGS and RBB acknowledge both PID2023-151567NB-I00 / 10.13039/501100011033 and RadioForegroundsPlus HORIZON-CL4-2023-SPACE-01 / GA 101135036 projects. RBB, LB and RGS thank the Red de Investigación RED2022-134715-T funded by MCIN/AEI/10.13039/501100011033.\\
The authors would like to faithfully thank the COSMOGLOBE Collaboration for making publicly available the sky map divided in regions that we used along this work.\\
This research has made use of the packages \texttt{Matplotlib} \citep{matplotlib}, \texttt{Keras} \citep{KER}, \texttt{Numpy} \citep{numpy}, \texttt{Namaster} \citep{NAMASTER}, \texttt{HEALPix} \citep{GOR05} and \texttt{Healpy} \citep{zon19} packages.
\end{acknowledgements}

\bibliographystyle{aa} 
\bibliography{main} 

\appendix

\section{Additional tests}
\label{sec:additional_tests}

In this Section, we describe additional methodological tests running the neural network trained on different maps with respect to the ones that formed the train dataset. In particular, Appendix \ref{sec:mode_loss} shows how the estimations of the network could vary due to the mode loss after the application of a filter of declination to the maps for removing RFI residual artifacts \citep{RUB23}. Appendix \ref{sec:curvature_models} covers how the estimations of the network when the synchrotron model used in the test data is a power-law plus a curvature term are consistent with the power-law without curvature model ones. Finally, Appendix \ref{sec:noise_complexity} explores how the noise between frequencies of the same horn correlations described in \cite{RUB23} could affect the estimations from the neural network presented in the main text.

\subsubsection{Mode loss due to the FDEC filtering}
\label{sec:mode_loss}

One of the methods applied during the post-processing of the QUIJOTE MFI data was the use of a declination-dependent function (hereafter FDEC) to correct for residual RFI signals appearing at fixed azimuth locations, as well as large-scale patterns \citep{RUB23}. Consequently, the resulting publicly available maps may contain some residual artifacts, which can affect our selected patches and the results obtained from fitting the power spectra, especially at low multipoles. Since we are fitting the power spectra from the multipole $l = 50$, we decided to not include this effect in the training set simulations. However, in this Section, we present a robustness test on wether this effect could led to increasing systematic uncertainties either in a particular region and/or generally.

Thus, we train again the neural network by using the same simulations as in the main text but after applying the FDEC filter, and we compare this network with the one trained without using the filter on the same test dataset, formed by simulations filtered by the FDEC. We then compute the power spectra of both cases and fit them as in the main text, estimating the relative errors between both cases for each synchrotron parameter.

Table \ref{tab:table_fdec} shows the relative error between both networks (trained with and without FDEC-filtered simulations) for each synchrotron parameter (the synchrotron amplitudes $A_{EE}$ and $A_{BB}$, the indices $\alpha_{EE}$ and $\alpha_{BB}$, and the amplitude ratio $A_{BB}/A_{EE}$).

First, the global behavior, both in Galactic (GP) and extragalactic (HLR) regions indicates that the application of the FDEC filter is slightly relevant for the B-mode parameters, with relative errors around 3.7\% for $A_{BB}$ in the GP regions and for HLR in the case of $\alpha_{BB}$. The average errors, however, decrease to a maximum of 2.04\% for $\alpha_{BB}$.

Furthermore, higher errors appear particularly in certain regions, especially for the B-mode and in low signal-to-noise ones, with errors nearly 10\% in 3, 4 and 8 regions for $A_{BB}$ and a relevant error of 18.31 for $\alpha_{BB}$ in region 2. Although these errors might not be overlooked, they possibly due to the poor sensitivity of QUIJOTE in such low SNR regions. 

Conversely, the North Polar Spur (Region 13) exhibits a remarkably high relative error of 13\% for $A_{BB}$ and 12.84\% for the $A_{BB}/A_{EE}$ ratio. Unlike the low SNR patches, the NPS is a highly significant and bright Galactic structure. This discrepancy cannot be attributed then to bad sensitivity instead, it represents a clear systematic effect introduced by the FDEC filter. This finding is consistent with previous analyses \citep{dlHOZ23}, which demonstrated that the FDEC filter significantly affects the recovery of the synchrotron spectral index $\beta_{S}$ in this region due to its large-scale spatial distribution. Consequently, we include a conservative systematic uncertainty of $\sim$ 13\% for the B-mode parameters exclusively in Region 13 to account for the residual effects of the FDEC post-processing routine.

Therefore, despite this region, we assume the higher errors from not applying the FDEC filter to the training dataset to the need of more sensible data, which will be further investigated in a future work when combining QUIJOTE with WMAP and \textit{Planck}.

\begin{table}[ht]
\caption{Relative error of the power-law synchrotron model parameters estimated by the network trained with FDEC-filtered maps and the one without filtering, both of them evaluated on the same FDEC-filtered maps at the sky regions defined in Table \ref{tab:table_regions}. The rows are the same than the ones explained in Table \ref{tab:table_observations}.}
\label{tab:table_fdec}
\centering
\footnotesize
\setlength{\tabcolsep}{4pt}
\begin{tabular}{cccccc}
\hline
Region & $A_{EE}$ (\%) & $A_{BB}$ (\%) & $\alpha_{EE}$ (\%) & $\alpha_{BB}$ (\%) & $A_{BB}/A_{EE}$ (\%) \\
\hline
1  & 4.66 & 4.54  & 0.58 & 1.63  & 8.79  \\
2  & 2.60 & 5.65  & 1.15 & 18.31 & 2.96  \\
3  & 1.13 & 8.73  & 0.54 & 7.19  & 9.97  \\
4  & 3.72 & 9.08  & 0.05 & 7.61  & 12.34 \\
6  & 0.70 & 0.19  & 0.97 & 6.31  & 0.88  \\
8  & 1.21 & 11.82 & 0.54 & 8.54  & 10.74 \\
9  & 1.46 & 6.87  & 1.55 & 3.36  & 8.45  \\
13 & 0.22 & 13.09 & 0.19 & 2.99  & 12.84 \\
17 & 0.20 & 4.34  & 0.61 & 0.06  & 4.14  \\
18 & 0.83 & 4.58  & 0.27 & 2.33  & 5.37  \\
19 & 0.62 & 3.32  & 0.52 & 6.16  & 3.92  \\
20 & 0.02 & 2.41  & 0.68 & 1.42  & 2.39  \\
\hline
GP  & 0.31 & 3.66  & 0.52 & 1.30  & 3.95  \\
HLR & 1.01 & 1.09  & 0.55 & 3.70  & 0.18  \\
All & 0.78 & 0.49  & 0.54 & 2.04  & 1.20  \\
\hline
\end{tabular}
\end{table}

\subsubsection{Testing the neural network with a power-law plus curvature synchrotron model}
\label{sec:curvature_models}

In this Section, we explore the performance of the neural network we have trained with a power-law synchrotron model, but tested on simulations with a synchrotron power-law plus a curvature term model. In particular, we use the s7 synchrotron model from PySM \citep{THO17}. Once the sky is simulated, we cut patches at the same positions described in Table \ref{tab:table_regions}.

\begin{figure*}[t]
\centering
\minipage{0.5\textwidth}
\includegraphics[width=\linewidth]{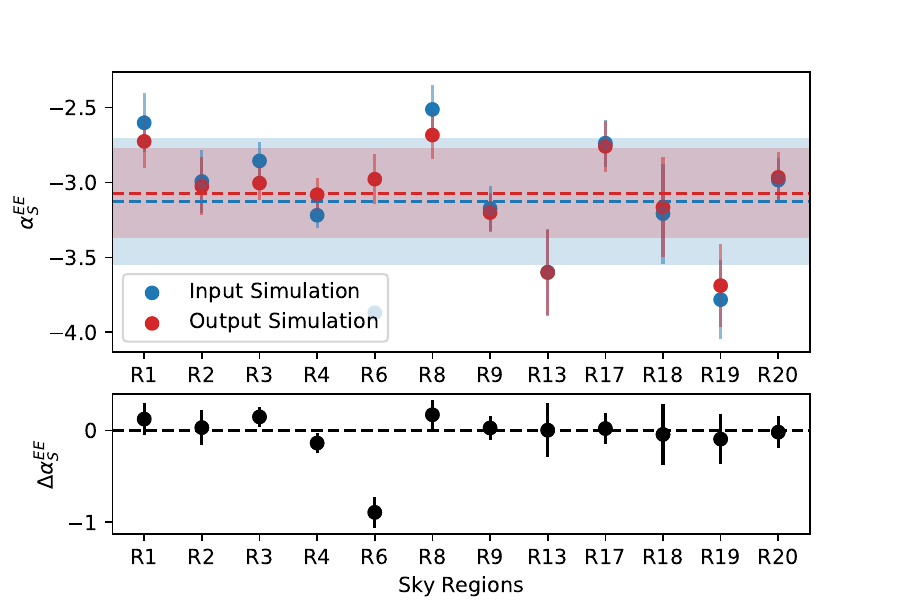}
\endminipage\hfill
\minipage{0.5\textwidth}%
  \includegraphics[width=\linewidth]{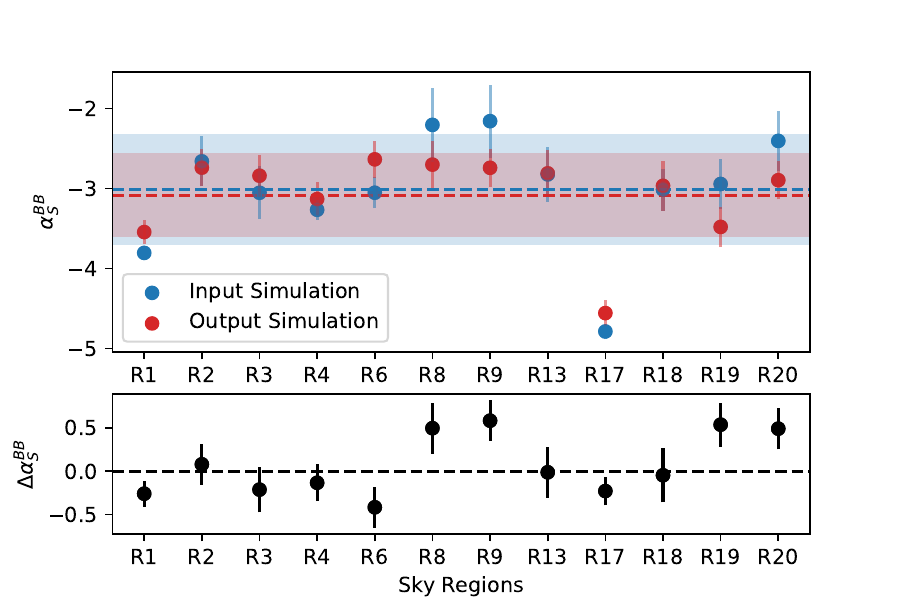}
\endminipage
\caption{Polarized synchrotron index $\alpha_{S}$ comparison at 11 GHz between the input s7 synchrotron model (blue dots) and the outputs from the neural network (in red), after fitting both $EE$ (left panel) and $BB$ spectra (right panel) for each sky region, while bottom subpanel displays the absolute error between them. The average index is shown in both panels as a dashed blue and red lines, respectively. The average difference is represented as a dashed black line. The errorbars for each point show the 1$\sigma$ uncertainty for each sky region evaluated over 100 simulations. Shaded colored areas show the 1$\sigma$ uncertainties for each average value.}
\label{Fig:Index_s7}
\end{figure*}

\begin{figure*}[t]
\centering
\minipage{0.5\textwidth}
\includegraphics[width=\linewidth]{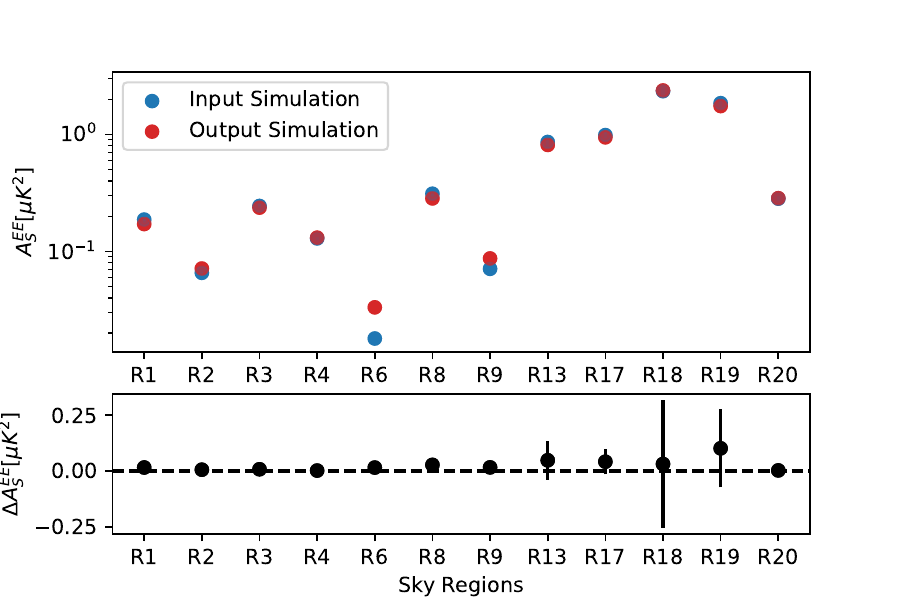}
\endminipage\hfill
\minipage{0.5\textwidth}%
  \includegraphics[width=\linewidth]{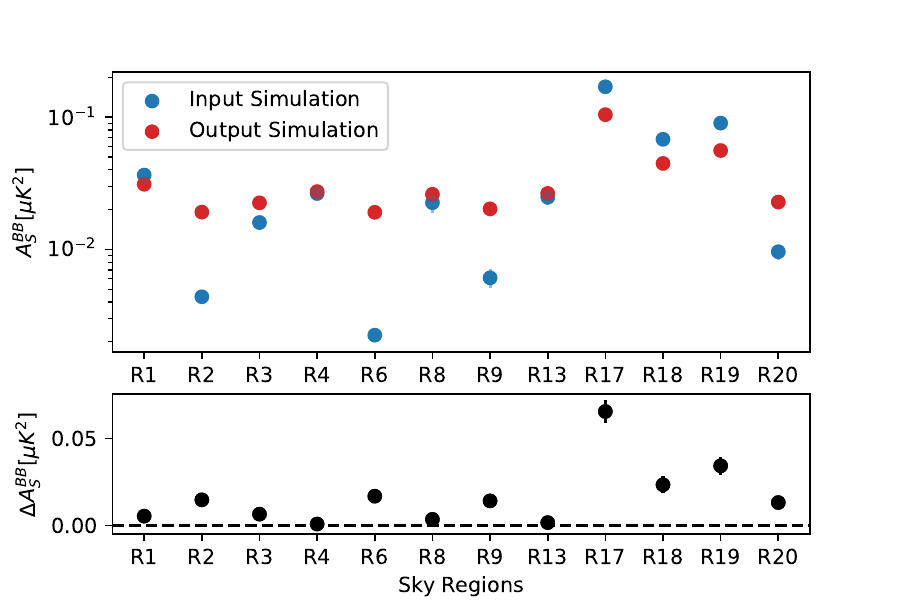}
\endminipage
\caption{Polarized synchrotron amplitude $A_{S}$ comparison at 11 GHz in logarithmic scale between the input s7 synchrotron model (blue dots) and the outputs from the neural network (in red), after fitting both $EE$ (left panel) and $BB$ spectra (right panel) for each sky region, while bottom subpanel displays in linear scale the absolute error between them. The average difference is represented as a dashed black line. The errorbars for each point show the 1$\sigma$ uncertainty for each sky region evaluated over 100 simulations. Shaded colored area show the 1$\sigma$ uncertainty for the absolute error average value.}
\label{Fig:Amplitude_s7}
\end{figure*}

\begin{figure}[ht]
\centering
\includegraphics[width=\linewidth]{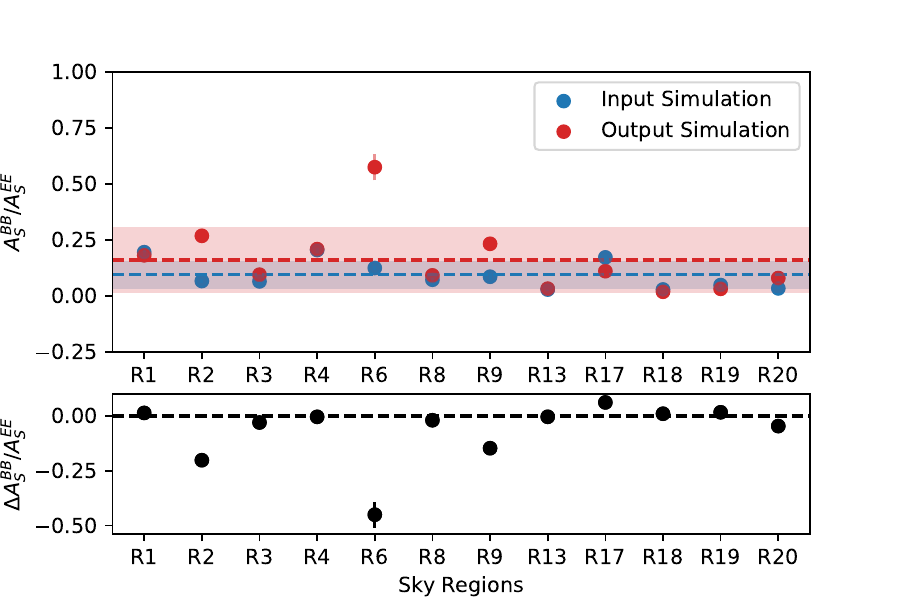}
\caption{Polarized synchrotron ratio comparison between $B$ and $E$ amplitudes at 11 GHz between the input s7 synchrotron model (blue dots) and the outputs from the neural network (in red), after fitting both spectra for each sky region. The bottom subpanel displays the absolute error between them. The average ratio is shown as a dashed blue and red lines, respectively. Shaded colored areas show the 1$\sigma$ respective uncertainties. The average difference is represented as a dashed black line. The errorbars for each point show the 1$\sigma$ uncertainty for each sky region propagated from $A_{S}^{BB}$ and $A_{S}^{EE}$ uncertainties using equation \ref{eq:error_propagation} evaluated over 100 simulations. Shaded colored area show the 1$\sigma$ uncertainty for the absolute error average value.}
\label{Fig:r_s7}
\end{figure}

The results are shown in Figures \ref{Fig:Index_s7}, \ref{Fig:Amplitude_s7}, and \ref{Fig:r_s7} for the indices $\alpha_{EE}$ and $\alpha_{BB}$, the amplitudes $A_{EE}$ and $A_{BB}$, and the amplitude ratio $A_{BB}/A_{EE}$, respectively. All the figures show the same metrics than the ones in Figures \ref{Fig:Simulations_SpectralIndex_ByRegion}, \ref{Fig:Simulations_Amplitude_ByRegion} and \ref{Fig:Simulations_r_ByRegion} for the synchrotron s5 model analysis.

We find that the results are consistent with, and very similar to, those obtained for the model without curvature, with the only noticeable difference being an improvement in the uncertainties of the $\alpha_{EE}$ and $\alpha_{BB}$ indices in the North Polar Spur region with respect to the s5 simulations. However, in most of the cases, we found that adding a curvature term, at least following this methodology, does not practically affect the uncertainties presented in the real data analysis.

Based on these results, we decided not to perform a new training using s7 simulations in the training dataset. However, future work combining QUIJOTE MFI and MFI2 data with \textit{Planck} and WMAP observations may reveal larger deviations in regions where curvature appears to be more significant, in particular in Regions 17 and 18 in this work, and especially in the southern hemisphere, as discussed in \cite{IRF25}.

\subsubsection{Noise correlations between frequencies of the same horn}
\label{sec:noise_complexity}

The QUIJOTE MFI maps, after the post-processing, still have several systematics that could affect scientific estimations from component separation methods. These systematics are extensively described in the QUIJOTE papers, especially in \cite{RUB23}, the noise correlation between the same horn being one of the most relevant in component separation analyzes. In this Section, we statistically study how much this systematics could affect the network estimations both totally and/or by region, being trained with only white isotropic noise, but tested both with such simplistic model and with the official QUIJOTE simulations, which include not only the 1$f$ term into the noise model but also instrumental systematics.

We evaluated the consistency of the reconstructed sky signal across independent realizations by comparing the recovered maps within each sky region, for both simplistic and complex noise cases. For every region, ten independent realizations of the reconstructed patch are available. The similarity between realizations was quantified using the Pearson correlation coefficient computed pixel-by-pixel.

For each region, the reconstructed maps were reshaped into one-dimensional vectors, and the correlation coefficient was calculated for all independent pairs of realizations, yielding 45 correlation values per region. The resulting correlations provide a direct measure of the reproducibility of the reconstructed sky signal across different realizations of the simulated data.

\begin{figure*}[t]
\centering
\minipage{0.5\textwidth}
\includegraphics[width=\linewidth]{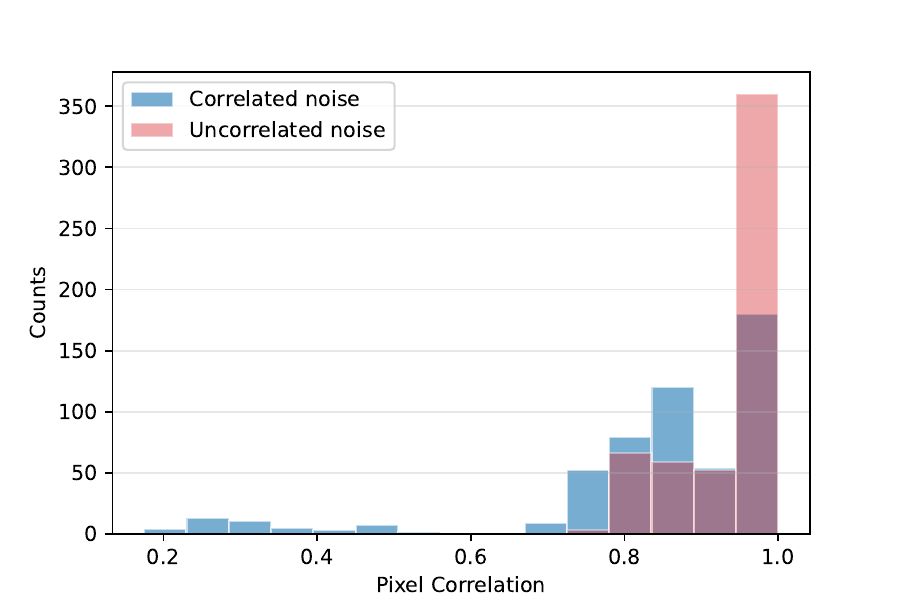}
\endminipage\hfill
\minipage{0.5\textwidth}%
  \includegraphics[width=\linewidth]{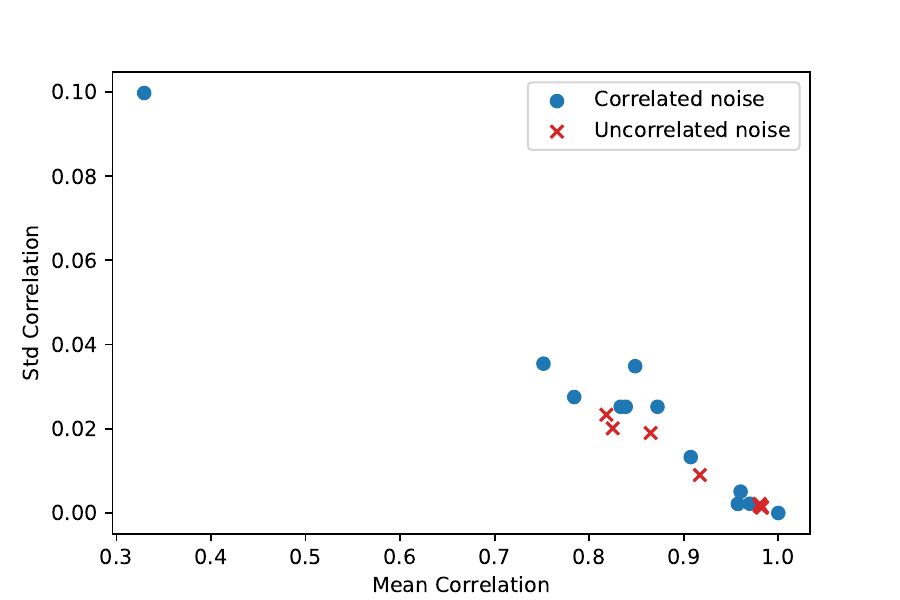}
\endminipage
\caption{Statistical recovery of several sky realizations using QUIJOTE noise simulations, with 1/$f$ term and noise correlations between frequencies of the same horn, which are called correlated noise (in blue) and compared with the uncorrelated one (in red) used along this work. In particular, the left panel shows an histogram combining the results from all sky regions defined in Table \ref{tab:table_regions} with the global distribution of the correlation coefficients between all the pair realizations. On the other hand, the right panel presents a scatter plot with the correlations at the region level comparing the mean correlation with respect to the standard deviation correlation for the correlated case (blue dots) and the uncorrelated case (red dots).}
\label{Fig:noise_correlations}
\end{figure*}

The global distribution of these correlation coefficients was then examined by constructing a histogram that combines the results of all regions for both correlated noise (which we called complex noise model) in blue and uncorrelated noise (called simplistic noise model) in red, which is shown in Figure \ref{Fig:noise_correlations}, left panel. This distribution summarizes the overall level of agreement between the reconstructed realizations and serves as a diagnostic of the stability of the reconstruction procedure. As shown, since most of the pixel correlations lie in the 0.8-1 interval, the recovery of the different sky realizations seems to be reliable for both kind of noise simulations. 

In addition, we characterized the statistical properties of the correlations at the region level by computing the mean correlation and its standard deviation for each region, which is shown in Figure \ref{Fig:noise_correlations}, right panel. These quantities were visualized in a scatter plot of mean correlation versus standard deviation, allowing a direct comparison of the reconstruction stability across different regions of the sky for the two noise models, each being represented by the same color as for the histogram. Regions with higher mean correlations and smaller dispersions correspond to more stable reconstructions across realizations, most of them in the interval of the mean correlation of 0.75-1, and with a small variance of 0-0.04. However, region 4 shows a totally opposite statistical behavior, probably because of its low signal-to-noise ratio. Despite for this region in particular, it seems that correlated noise in the training dataset does not affect the results of the neural network.

\section{Component separation model based on fully-convolutional neural networks}
\label{sec:model}

On the one hand, our approach is grounded in the physical principles described in \cite{ERI06} and \cite{ERI08}. On the other hand, it adheres to the theoretical framework of CNNs and FCNs, as detailed in \cite{Rum86}, \cite{LeC98}, and \cite{GOO16}. 

Our model operates on patches $x_{\nu}$ of the microwave sky at various frequencies $\nu$, each modeled as a linear mixture of $c$ astrophysical sources. These sources are represented by emission maps $e_{\nu}$, convolved with the instrumental beam of the experiment $i_{\nu}$ and superimposed with instrumental noise $n_{\nu}$:
\begin{equation}
    x_{\nu} = i_{\nu} * e_{\nu} + n_{\nu},
\label{eq:model}
\end{equation}
where $*$ denotes convolution. The emission maps are composed as a linear combination of foreground components, each with distinct spectral properties:
\begin{equation}
    e_{\nu} = \sum_{c} A_{c, \nu} \, s_{c},
\label{eq:emission_maps}
\end{equation}
with $A$ denoting the mixing matrix and $s$ representing the set of component maps. Substituting into Equation \eqref{eq:model}, the full model becomes:
\begin{equation}
    x_{\nu} = i_{\nu} * \left( \sum_{c} A_{c, \nu} \, s_{c} \right) + n_{\nu}.
\label{eq:final_model}
\end{equation}

Initially, all model parameters are set randomly. The neural network ingests a tensor of shape $N_{\text{pix}} \times N_{\text{pix}} \times N_{\nu}$, where $N_{\text{pix}}=256$ and $N_{\nu}$ denote the number of frequency channels. Spectral features are captured via multidimensional strided convolutions \citep{GOO10}:
\begin{equation}
    H_{\nu, i, j} = \sum_{k, m} W_{k, m, i} \, V_{s \circ \nu + k, m, j},
\label{eq:convolutional_blocks}
\end{equation}
where $H$ are the hidden units, $V$ are the input values and $W$ is the convolutional kernel. The operator $s$ represents the stride vector, and $\circ$ indicates elemental multiplication.

Subsequent convolutional blocks apply similar operations, utilizing the output from previous layers as input.

In the decoder (deconvolutional blocks), we perform a strided transposed convolution across spectral dimensions:
\begin{equation}
    R_{q, m, j} = \sum_{\nu, k \,|\, s \circ \nu + k = q} \sum_{i} W_{k, m, i} \, H_{\nu, i, j},
\label{eq:deconvolutional_blocks}
\end{equation}
where $R$ denotes the deconvolved output and $H$ the activations of the previous layer. The condition after the vertical bar applies a modulo constraint.

The kernels $W$ across all convolutional and deconvolutional layers are updated each epoch by mnimizing the mean squared error:
\begin{equation}
    MSE = \frac{1}{2} \left| y - y' \right|^2,
\label{eq:mse}
\end{equation}
where $y$ is the predicted output and $y'$ is the target (i.e., the true CMB map). The computed loss is then backpropagated to update the layer parameters.

In the final deconvolutional layer, all spectral data is projected into a single channel, producing a reconstructed map at the reference frequency $\nu_{0} = 217$ GHz:
\begin{equation}
    \Tilde{x}_{\nu_{0}} = i_{\nu_{0}} * e_{\text{CMB}, \nu_{0}} + \Tilde{n},
\label{eq:final_map}
\end{equation}
where $\Tilde{n}$ represents deconvolution-induced noise and $e_{\text{CMB}, \nu_{0}}$ is the CMB component.

The model parameters are updated each epoch according to:
\begin{equation}
    \theta_{t+1} = \theta_{t} + \Delta \theta_{t},
\label{eq:parameter}
\end{equation}
with the update step defined as:
\begin{equation}
    \Delta \theta_{t} = - \alpha \, g_{t},
\label{eq:parameter_variation}
\end{equation}
where $g_{t}$ is the gradient and $\alpha$ the learning rate, which is computed using the AdaGrad algorithm \citep{Duchi2011}:
\begin{equation}
    \alpha = \frac{\eta}{\sqrt{G_{t} + \epsilon}},
\label{eq:adagrad}
\end{equation}
with $G_{t}$ being the accumulated squared gradients, $\epsilon$ a stability term, and $\eta = 0.05$.

Upon propagating all minibatches through the network, the final gradient is computed at the last deconvolutional block as:
\begin{equation}
    g_{t} = \frac{\partial E}{\partial O_{\nu_{0}}} = \frac{\partial}{\partial O_{\nu_{0}}} \left( \frac{1}{2} | O_{\nu_{0}} - l_{\nu_{0}} |^2 \right) = | O_{\nu_{0}} - l_{\nu_{0}} |,
\label{eq:last_deconvolutional}
\end{equation}
where $l_{\nu_{0}}$ is the ground truth label (CMB signal).

The backpropagation then proceeds by applying the chain rule. For the previous deconvolutional layer:
\begin{equation}
\begin{split}
    g_{t, W} &= \frac{\partial O_{\nu_{0}, LD}}{\partial W_{LD - 1}} 
    = \frac{\partial O_{\nu_{0}, LD}}{\partial O_{LD - 1}} \cdot \frac{\partial O_{LD - 1}}{\partial W_{LD - 1}} \\
    &= g_{t, LD} \cdot f_{LD - 1},
\end{split}
\label{eq:gradients1}
\end{equation}
\begin{equation}
\begin{split}
    g_{t, f} &= \frac{\partial O_{\nu_{0}, LD}}{\partial f_{LD - 1}} 
    = \frac{\partial O_{\nu_{0}, LD}}{\partial O_{LD - 1}} \cdot \frac{\partial O_{LD - 1}}{\partial f_{LD - 1}} \\
    &= g_{t, LD} \cdot W_{LD - 1},
\end{split}
\label{eq:gradients2}
\end{equation}
where $g_{t, W}$ and $g_{t, f}$ are the gradients with respect to the weights and filters. These are derived using the 2D convolution formula:
\begin{equation}
    O_{i, j} = \sum_{m = 0}^{M} \sum_{n = 0}^{N} f(i - m, j - n) \, W(m, n).
\label{eq:convolution_matrix}
\end{equation}

Here, LD and LD$-1$ represent the last and penultimate deconvolutional blocks. This approach is applied similarly  to prior deconvolutional layers.

Analogously, convolutional layers use the following:
\begin{equation}
    g_{t, W} = g_{t, FD} \cdot f_{LC},
\label{eq:gradients_convolutional1}
\end{equation}
\begin{equation}
    g_{t, f} = g_{t, FD} \cdot W_{LC},
\label{eq:gradients_convolutional2}
\end{equation}
where FD and LC refer to the first deconvolutional and last convolutional blocks, respectively. All other convolutional blocks follow the same gradient-based update rules.

\section{Full sky maps}
\label{sec:full_sky_maps}

\begin{figure*}[t]
\centering
\minipage{0.5\textwidth}
\includegraphics[width=\linewidth]{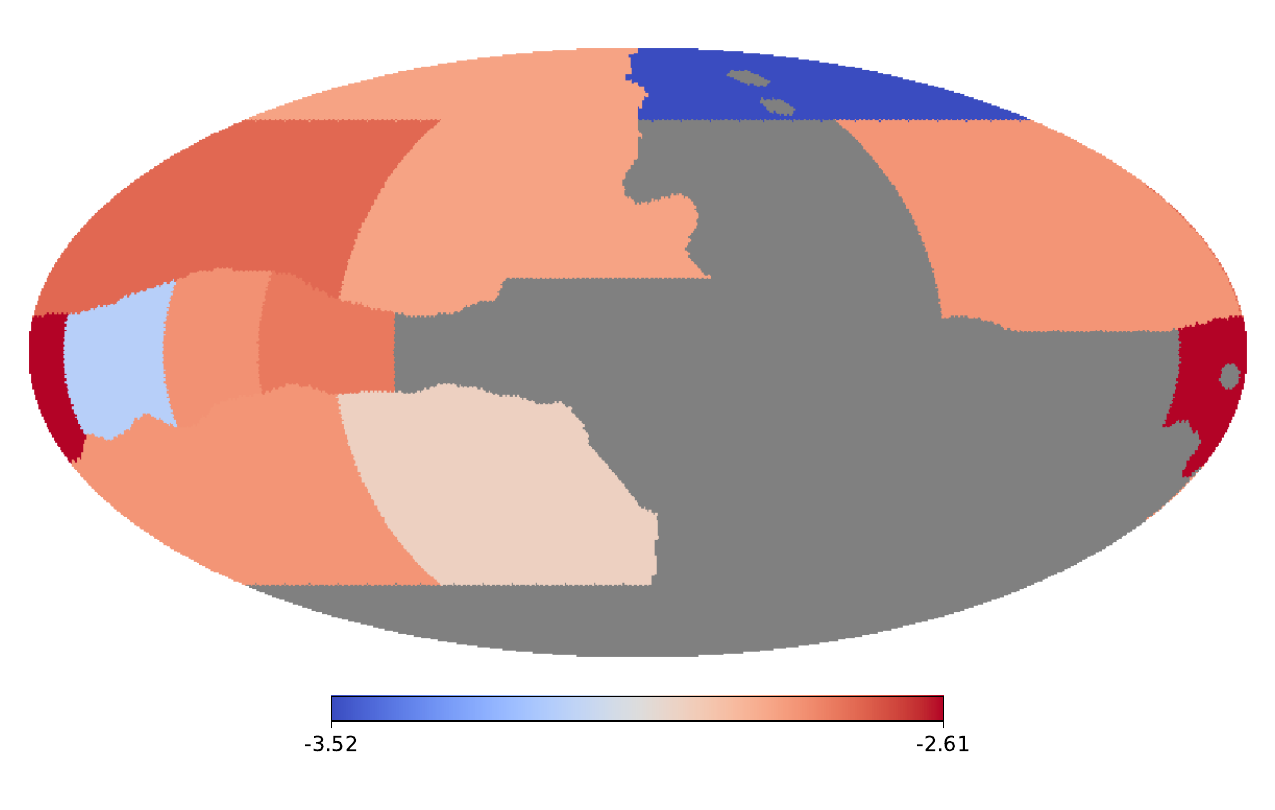}
\includegraphics[width=\linewidth]{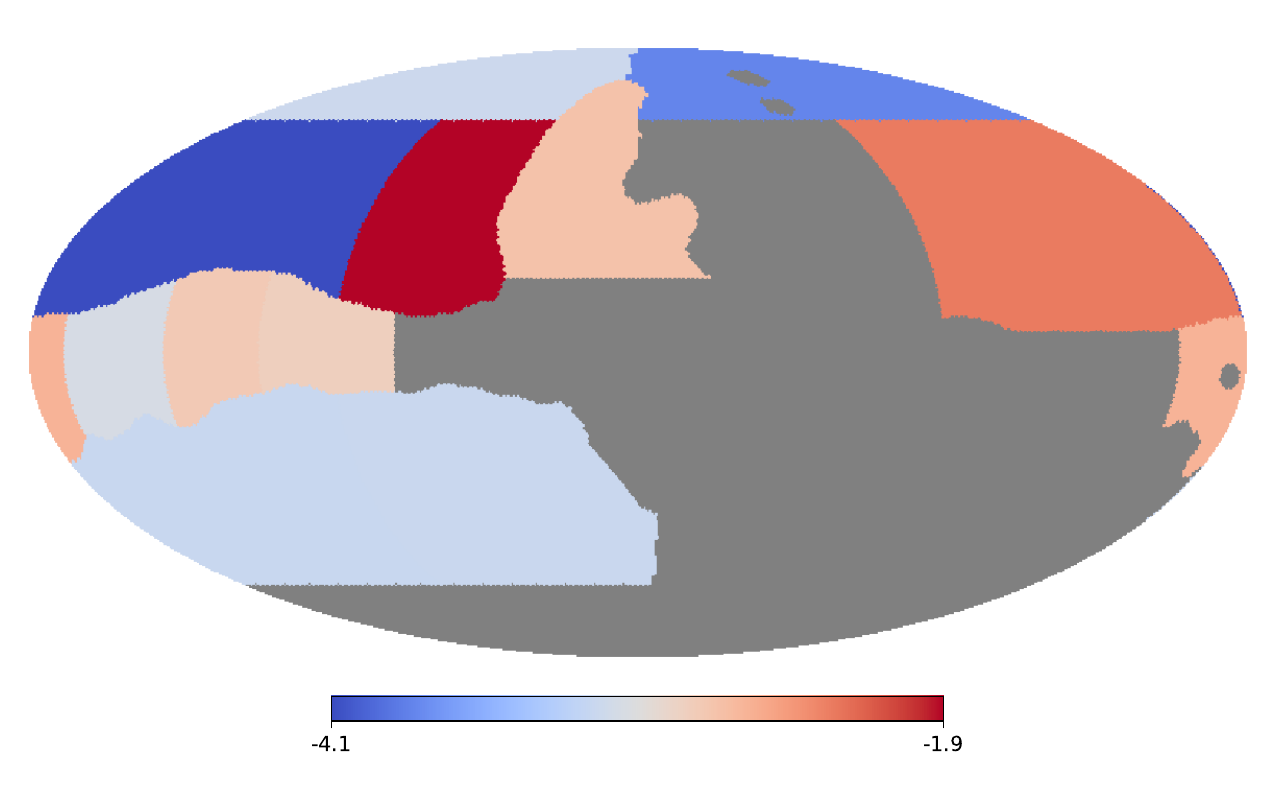}
\endminipage\hfill
\minipage{0.5\textwidth}%
  \includegraphics[width=\linewidth]{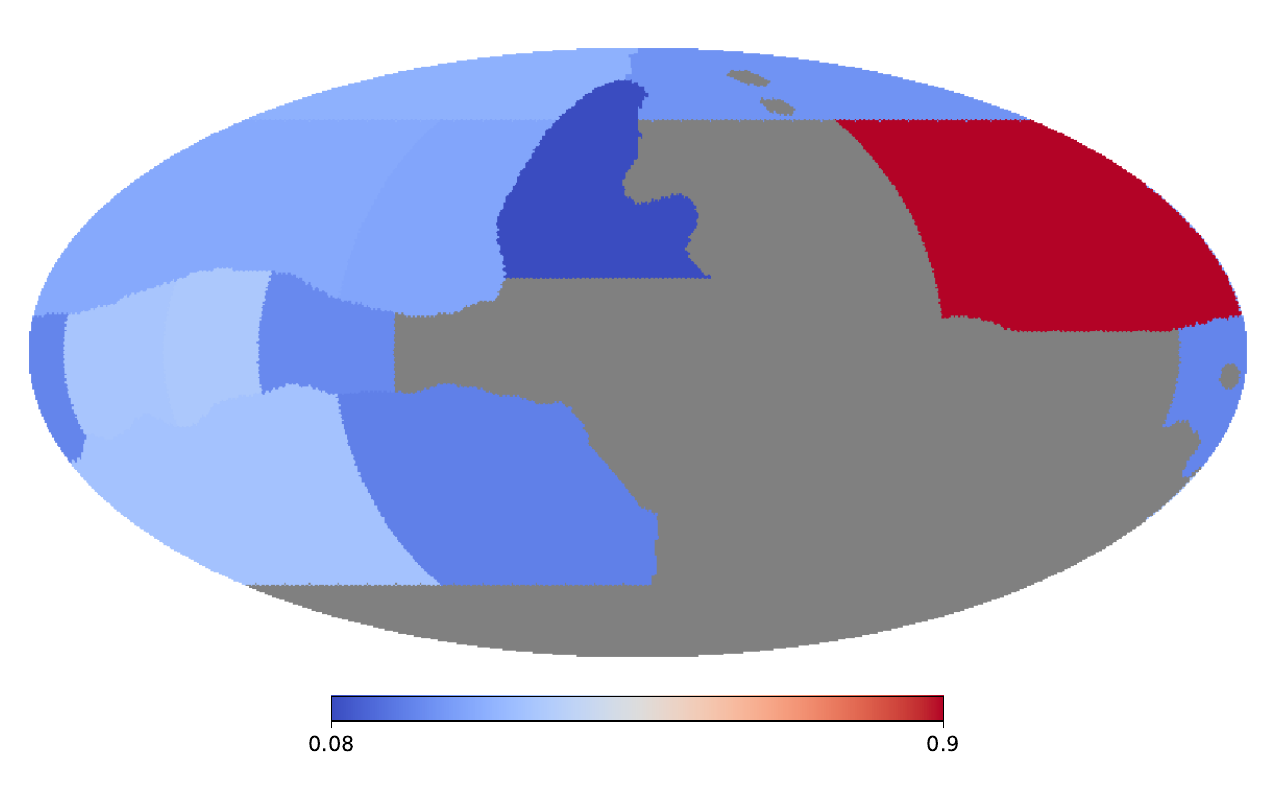}
  \includegraphics[width=\linewidth]{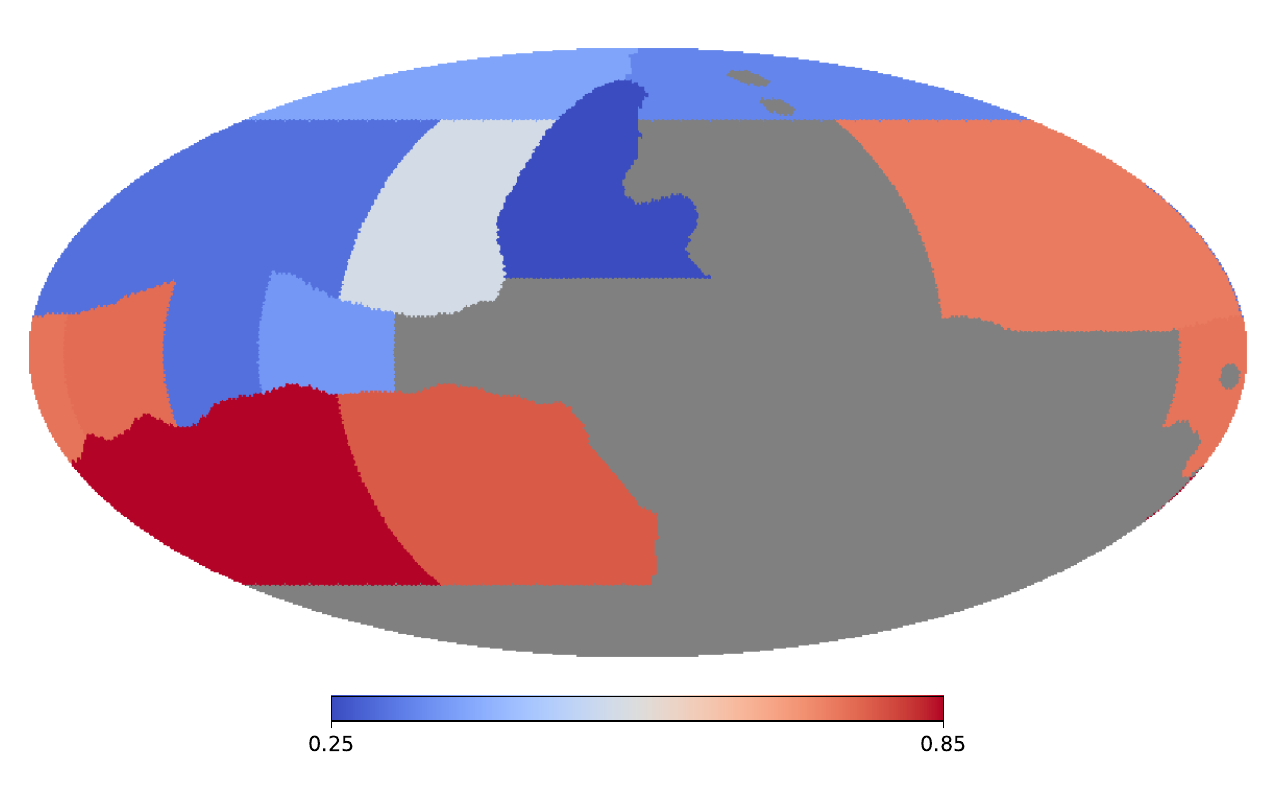}
\endminipage
\caption{Mollview projection of our estimates for the $\alpha_{S}^{EE}$ (top pannel) and $\alpha_{S}^{BB}$ indeces in all the regions covered by the QUIJOTE MFI instrument at 11 GHz. The relative uncertainties (statistical plus systematic) for each region are shown in the right panel. The regions that we cannot analyze due to the QUIJOTE MFI covering are masked out in grey areas.}
\label{Fig:all_sky_alpha}
\end{figure*}

\begin{figure*}[t]
\centering
\minipage{0.5\textwidth}
\includegraphics[width=\linewidth]{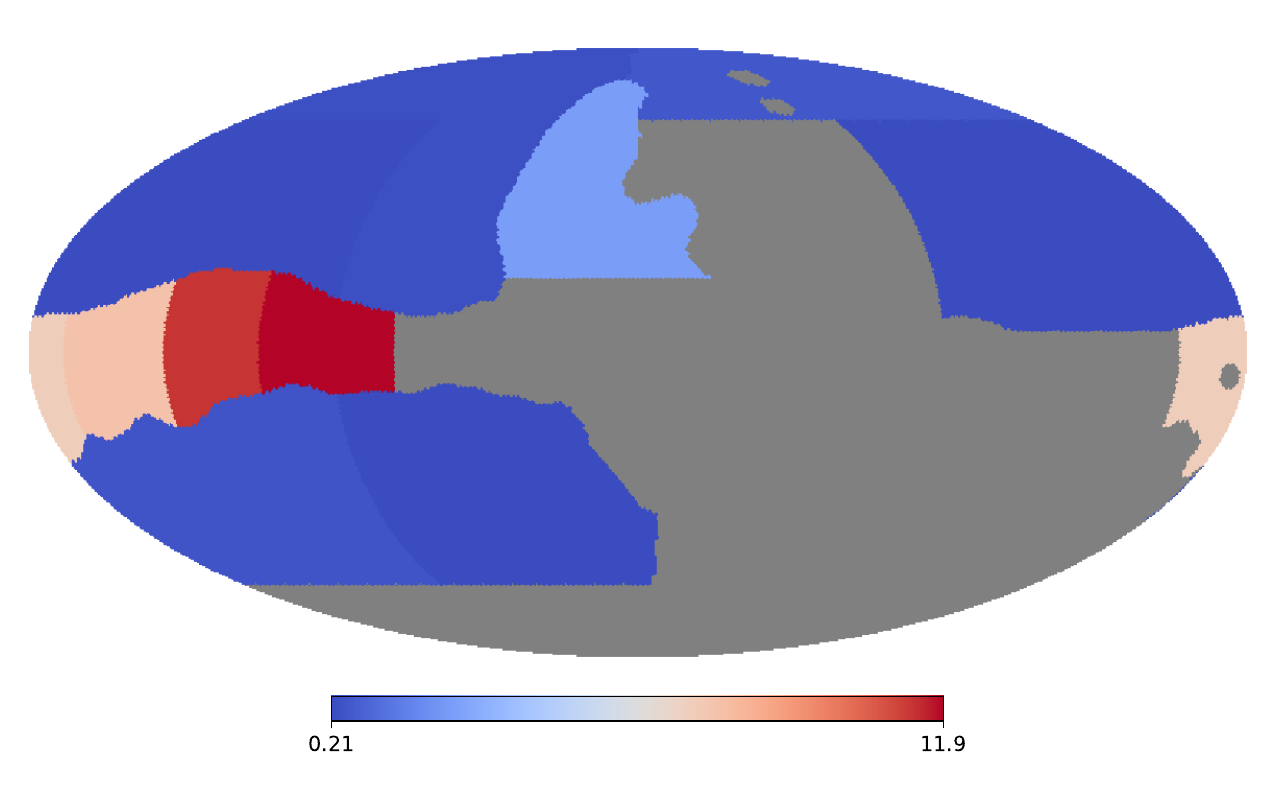}
\includegraphics[width=\linewidth]{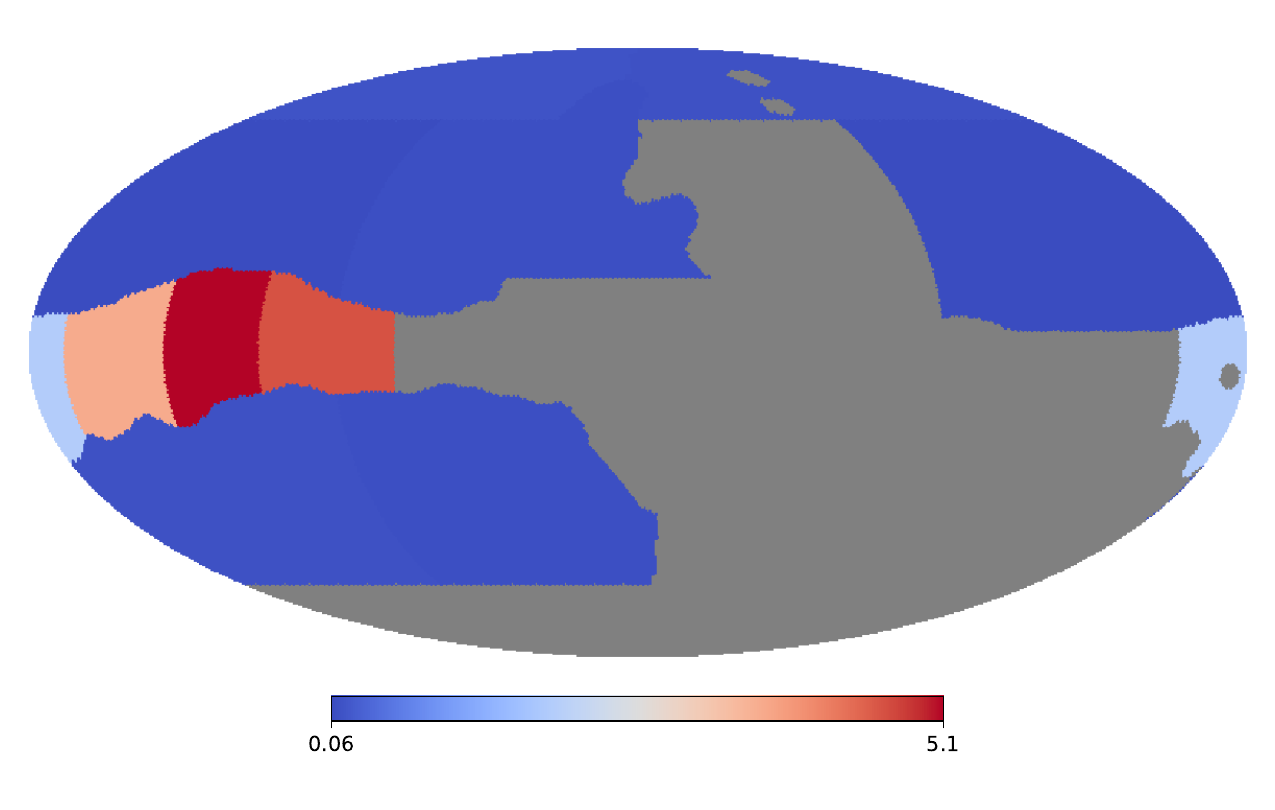}
\endminipage\hfill
\minipage{0.5\textwidth}%
  \includegraphics[width=\linewidth]{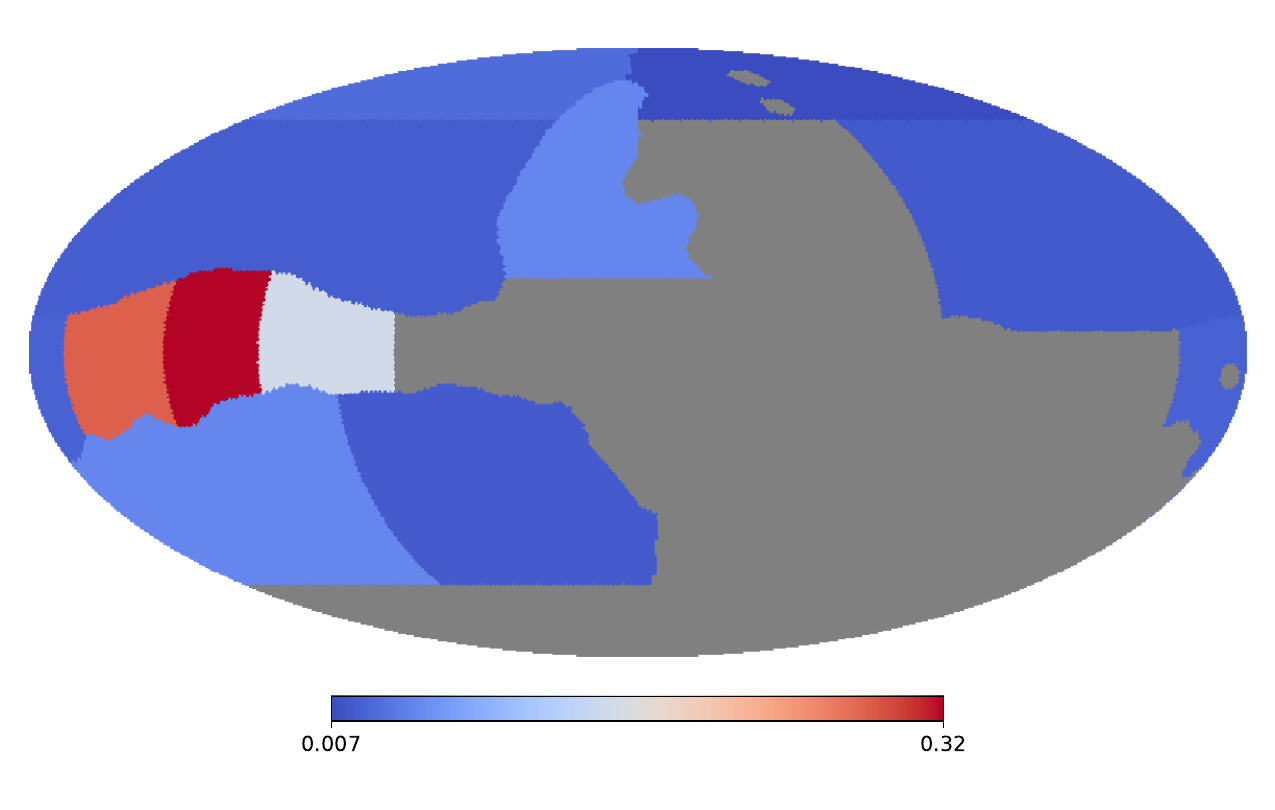}
  \includegraphics[width=\linewidth]{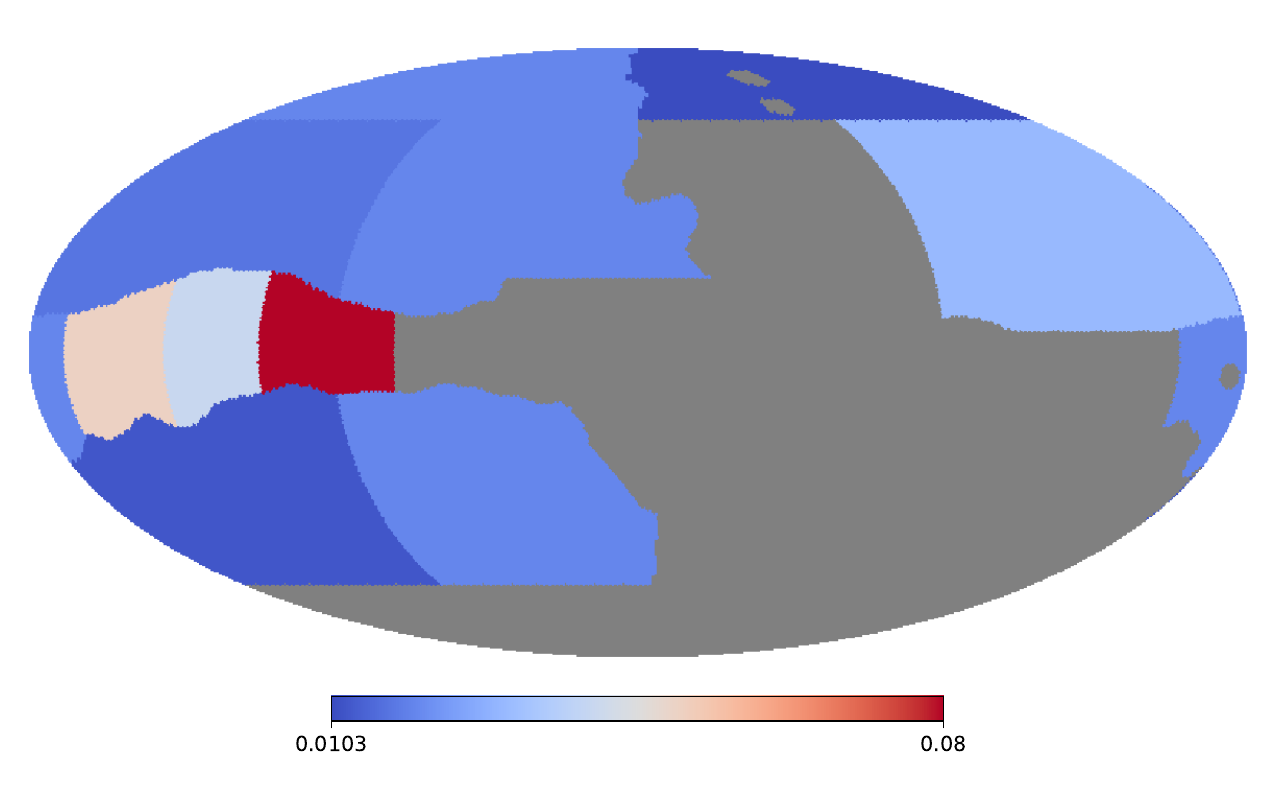}
\endminipage
\caption{Mollview projection of our estimates for the $A_{S}^{EE}$ (top pannel) and $A_{S}^{BB}$ indeces in all the regions covered by the QUIJOTE MFI instrument at 11 GHz. The relative uncertainties (statistical plus systematic) for each region are shown in the right panel. The regions that we cannot analyze due to the QUIJOTE MFI covering are masked out in grey areas.}
\label{Fig:all_sky_A}
\end{figure*}

\begin{figure*}[t]
\centering
\minipage{0.5\textwidth}
\includegraphics[width=\linewidth]{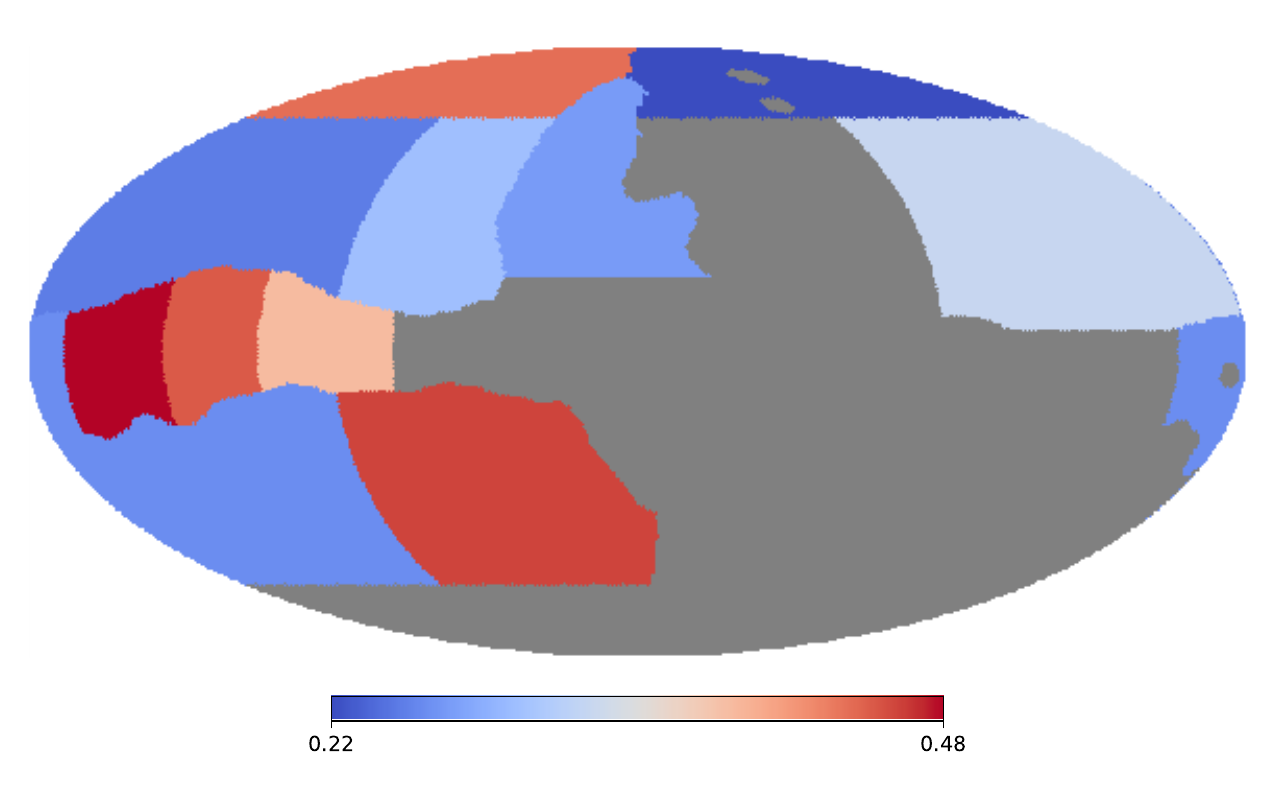}
\endminipage\hfill
\minipage{0.5\textwidth}%
  \includegraphics[width=\linewidth]{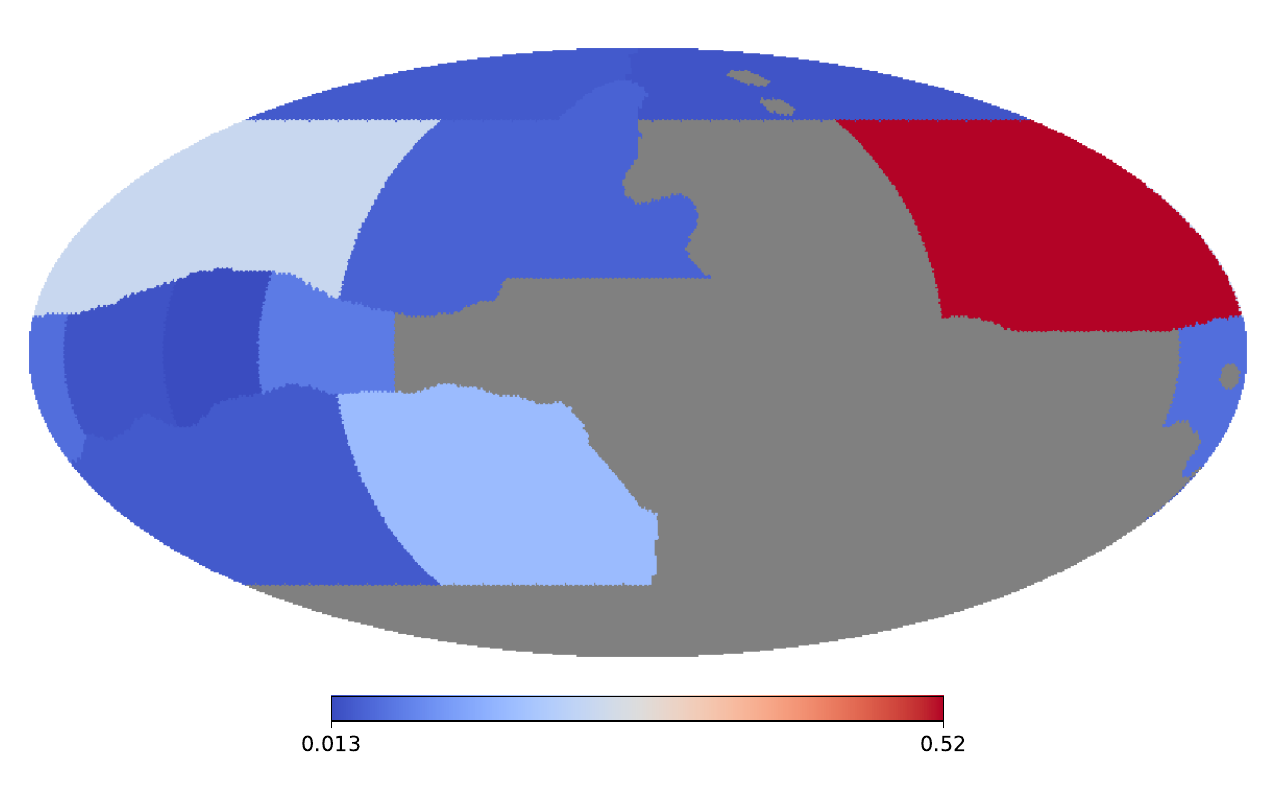}
\endminipage
\caption{Mollview projection of our estimates for the ratio between amplitudes $A_{S}^{BB}/A_{S}^{EE}$ in all the regions covered by the QUIJOTE MFI instrument at 11 GHz. Left panel: mean index for each region. The relative uncertainty (statistical plus systematical) for each region is shown in the right panel. The regions that we cannot analyze due to the QUIJOTE MFI covering are masked out in grey areas.}
\label{Fig:all_sky_ABtoAE}
\end{figure*}

In this appendix, we aim to show the same results as the ones presented in Section \ref{sec:results}, but displayed in full-sky maps and divided by the analyzed sky regions. Figure \ref{Fig:all_sky_alpha} shows the mollview projection of our estimates for the $\alpha_{S}$ index for both $E$ and $B$. Figure \ref{Fig:all_sky_A} displays the same but for the amplitude. Finally, Figure \ref{Fig:all_sky_ABtoAE} presents the same but for the ratio between the $B$-to-$E$ amplitudes.

\end{document}